\documentclass[fleqn,usenatbib]{mnras}

\usepackage{amsmath}
\usepackage{amssymb}
\usepackage{color}
\usepackage{epic}
\usepackage{epstopdf}
\usepackage{graphicx}
\usepackage{hyperref}
\usepackage{multirow}
\usepackage{natbib}
\usepackage{overpic}

\usepackage{color}
\AtBeginShipout{%
 \ifnum\value{page}>1 %
 \typeout{* Additional boxing of page `\thepage'}%
 \setbox\AtBeginShipoutBox=\hbox{\copy\AtBeginShipoutBox}%
 \fi
}

\title[RAiSE X: searching for radio galaxies in X-ray surveys]{RAiSE X: searching for radio galaxies in X-ray surveys}

\author[R. J. Turner et al.]{
Ross J. Turner$^{1}$\thanks{Email: turner.rj@icloud.com} and Stanislav S. Shabala$^{1,2}$\\
$^{1}$School of Natural Sciences, University of Tasmania, Private Bag 37, Hobart, 7001, Australia\\
$^{2}$ARC Centre of Excellence for All-Sky Astrophysics in 3 Dimensions (ASTRO 3D)}

\date{Accepted 2020 March 06. Received 2020 March 05; in original form 2019 December 12}

\pubyear{2019}

\begin{document}

\label{firstpage}
\pagerange{\pageref{firstpage}--\pageref{lastpage}}
\maketitle

\begin{abstract}

We model the X-ray surface brightness distribution of emission associated with {Fanaroff \& Riley} type-II radio galaxies. Our approach builds on the RAiSE dynamical model which describes broadband radio-frequency synchrotron evolution of jet-inflated lobes in a wide range of environments. The X-ray version of the model presented here includes: (1) inverse-Compton upscattering of cosmic microwave background radiation; (2) the dynamics of the shocked gas shell and associated bremsstrahlung radiation; and (3) emission from the surrounding ambient medium. We construct X-ray surface brightness maps for a mock catalogue of extended FR-IIs based on the technical characteristics of the \emph{eRosita} telescope. The integrated X-ray luminosity function at low redshifts ($z\leqslant1$) is found to strongly correlate with the density of the ambient medium in all but the most energetic sources, whilst at high-redshift ($z>1$) the majority of objects are dominated by inverse-Compton lobe emission due to the stronger cosmic microwave background radiation. By inspecting our mock spatial brightness distributions, we conclude that any extended X-ray detection can be attributed to AGN activity at redshifts $z\geqslant1$. We compare the expected detection rates of active and remnant high-redshift radio AGNs for \emph{eRosita} and \emph{LOFAR}, and future more sensitive surveys. We find {that} a factor of ten more remnants can be detected using X-ray wavelengths over radio frequencies at $z>2.2$, increasing to a factor of 100 for redshifts $z>3.1$.

\end{abstract}


\section{INTRODUCTION}
\label{sec:INTRODUCTION}

Most galaxies are known to harbour a supermassive black hole (SMBH) at their centre \citep{Magorrian+1998}, with the galaxy and SMBH growth rates strongly intertwined over their evolutionary history \citep{McNamara+2012}. 
The activity of the black hole is tied to the state of its surroundings, since, as hot cluster gas cools by bremsstrahlung radiation, it forms cooling flows that sink towards and are accreted by the SMBH, switching it on \citep{McNamara+2007}. This active galactic nucleus (AGN) imparts a fraction of its accreted energy back into its host environment through either radiative or kinetic-mode feedback \citep{Fabian+2012}. The radio lobes inflated by the relativistic jets of pair-plasma, which emanate from the accretion disk in the kinetic-mode, are responsible for pushing out vast amounts of gas from the host galaxy \citep{Nesvadba+2008, Morganti+2013} and thus suppress star formation \citep{Page+2012}, or sometimes trigger regions of enhanced growth along shock fronts \citep{Gaibler+2012}. \citet{Shankar+2009} find the growth of black holes closely tracks the star formation history of galaxies across cosmic time. The energy input by AGNs into their surroundings through kinetic feedback (primarily by shocks and pressure-volume work) prevent catastrophic cooling of the hot cluster gas \citep{Fabian+2003, Forman+2005, Mittal+2009}. Kinetic-mode feedback is also invoked in cosmological galaxy formation models to explain the missing stellar mass in the brightest galaxies towards the present epoch \citep{Croton+2006, Bower+2006, Vogelsberger+2014, Raouf+2017}.


The quantification of the energetics of kinetic-mode AGN feedback requires knowledge of the energy imparted in each outburst and the duty cycle of the black hole activity. Jet kinetic powers can be directly measured at radio-frequencies using techniques including hotspot luminosities \citep{Godfrey+2013} and spatial shifts of radio cores in VLBI images \citep{Konigl+1981, Lobanov+1998, Shabala+2012}, or more readily constrained in large sky surveys using dynamical models for the lobes populated by shock-accelerated synchrotron-emitting electrons \citep{Turner+2015, Turner+2018b, Hardcastle+2019}. Recent high sensitivity and resolution observations by the \emph{Low Frequency Array} (LOFAR) at low frequencies ($150\rm\, MHz$) have reinvigorated studies of the AGN duty cycle.  \citet{Sabater+2019} found {that} up to 100 per cent of all large galaxies host an active nucleus, noting that their accretion rate is modulated such that half the energy output is released in outbursts of increased activity lasting less than two percent of the time. Studies of individual objects and large sky surveys have also found populations of remnant \citep{Brienza+2017, Mahatma+2018} and restarted \citep{Mahatma+2019} radio galaxies, setting a lower bound for the duty-cycle. \citet{Turner+2018} used a radio source dynamical model to provide an upper limit on the duty cycle of $\delta < 0.15$ in the remnant B2\,0924+30. \citet{Shabala+2019} tighten the range of plausible duty cycles by comparing the size and luminosity functions of radio sources in the Lockman Hole to simulated functions based on dynamical models for a range of lifetime distributions.


The shock-accelerated electrons comprising extended radio AGN lobes not only emit synchrotron radiation at radio frequencies, but also produce X-ray emission at keV energies due to the inverse-Compton upscattering of cosmic microwave background photons by lower energy electrons. The inverse-Compton and synchrotron radiative loss mechanisms more rapidly deplete the higher energy ($\gamma \sim 10^4$) electrons involved in the synchrotron radiation compared to the lower energy ($\gamma \sim 10^3$) electrons responsible for the X-ray inverse-Compton emission \citep{Nath+2010, Mocz+2011}. Remnant radio galaxies are therefore expected to appear as inverse-Compton `ghosts' for some period of time after the cessation of jet activity before becoming undetectable at both radio and X-ray wavelengths \citep{BR+1999}. In particular, the extended X-ray source HDF\,130 at $z = 1.99$ has been identified as a remnant radio galaxy despite its double-lobed structure only being visible at X-ray wavelengths \citep{Fabian+2009}. Further inverse-Compton `ghosts' should be readily detectable at high redshift since the energy density of the cosmic microwave background increases as $(1 + z)^4$, offsettting the reduced X-ray flux density at greater distance \citep{Mocz+2011, Ghisellini+2014}. Conversely, only the youngest radio galaxies at high-redshift will be detectable  in radio observations due to the extreme radiative losses resulting from the strong microwave background energy density. \citet{Fabian+2014} investigate two distant quasars, ULAS J112001.48+064124.3 at $z = 7.1$ and SDSS J1030+0524 at $z = 6.3$, finding that powerful jets fuelled by super-Eddington accretion rates could exist but be undetectable with current surveys. However, the benefit of increased brightness at higher redshifts may be offset by the convolution of the extended and core emission, which can be of comparable magnitude to the brightest inverse-Compton lobes \citep{Mingo+2014}.

In this work, we extend the \emph{Radio AGNs in Semi-analytic Environments} \citep[RAiSE;][]{Turner+2015, Turner+2018a, Turner+2018} model for the dynamical evolution and synchrotron emissivity of active, remnant and restarted radio galaxies to X-ray wavelengths; in particular, we seek to quantify which radio galaxies have non-core associated X-ray emission detectable with the surface brightness sensitivity of current surveys (e.g. using \emph{Chandra} and \emph{eRosita}). Previous iterations of RAiSE have found success in: (1) reproducing surface brightness and spectral age maps for canonical FR-I (3C31) and FR-II (3C436) type sources \citep{Turner+2018a}; (2) deriving jet kinetic powers consistent with X-ray inverse-Compton measurements \citep{Turner+2018b}; and (3) accurately constraining the Hubble constant using low-redshift AGNs \citep{Turner+2019}. In this paper, we first extend RAiSE to calculate the integrated X-ray luminosity due to the inverse-Compton upscattering of cosmic microwave background radiation by lobe electrons, then create model X-ray surface brightness maps for lobes of \citet[FR;][]{FR+1974} type-II morphology (Section \ref{sec:LUMINOSITY MODEL}). These X-ray brightness maps are completed in Section \ref{sec:SHOCKED SHELL BREMSSTRAHLUNG RADIATION} by including the bremsstrahlung radiation both from the shocked shell of gas swept up between the bow shock and lobe plasma, and from the ambient medium. In Section \ref{sec:X-ray surface brightness maps}, we investigate the changing importance of the inverse-Compton and bremsstrahlung radiative mechanisms with redshift and intrinsic properties of the source. Finally, we create a mock catalogue of extended radio galaxies based on an observationally informed set of parameters; this catalogue is used to generate the X-ray luminosity function for FR-IIs over cosmic time, and characterise the effectiveness of both radio-frequency (\emph{SKA}-pathfinders) and X-ray surveys at detecting high-redshift remnants (Section \ref{sec:MOCK EXTENDED AGN POPULATION}).

The $\Lambda \rm CDM$ concordance cosmology with $\Omega_{\rm M} = 0.3$, $\Omega_\Lambda = 0.7$ and $H_0 = 70 \rm\,km \,s^{-1} \,Mpc^{-1}$ \citep{Planck+2016} is assumed throughout the paper.

\section{LOBE INVERSE-COMPTON EMISSIVITY}
\label{sec:LUMINOSITY MODEL}


The inverse-Compton emissivity model developed in this work is based on the synchrotron model presented by \citet{Turner+2018a}; it can therefore similarly be applied to any AGN dynamical model, be it analytical or a hydrodynamical simulation. The \citet{Turner+2018a} model assumes that the radiative losses are not taken into account self-consistently in the evolution of the lobe pressure. \citet{Hardcastle+2018} confirms that synchrotron radiation comprises less than ten percent of the input power for typical radio galaxies (at $z=0$), whilst inverse-Compton and bremsstrahlung radiation contribute even less to the energy loss during the source expansion. However, the cosmic microwave background radiation responsible for the inverse-Compton losses increases with redshift as $(1 + z)^4$; the inverse-Compton radiation can reach up to 40\% of the input power in a $100\rm\, Myr$ source at $z=1$, or all of the input power for a $10\rm\, Myr$ source at $z=4$. The results in this work do not consider objects beyond these extremes in age and redshift. The adiabatic expansion of the lobe is therefore assumed to take up the bulk of the input energy as explicitly considered in numerical and published analytical dynamical models.

\citet{Hardcastle+1998} proposed a technique to derive the inverse-Compton emissivity from the hotspots of powerful FR-IIs by integrating over the full electron and photon distribution. \citet{Nath+2010} and \citet{Mocz+2011} later both derived models for the inverse-Compton emission from radio lobes, with the formalisms consistent to those found in the dynamical and synchrotron emissivity models of \citet{KA+1997} and \citet{KDA+1997}. These two models make use of the electron energy distribution assumed in the synchrotron emissivity model, and will therefore have the same strengths and shortcomings as the calculation of the radio emissivity.

\subsection{Unresolved, continuously injected electron model}

The synchrotron-emitting electrons in radio sources have significant kinetic energy when compared to the energy of the cosmic microwave background (CMB) photons, and thus the electron energy can be transferred to a CMB photon through inverse-Compton scattering. 

Assuming that the electrons emit synchrotron radiation only at some critical frequency $\nu_{\rm syn} = \gamma^2 \nu_{\rm L}$, where $\nu_{\rm L}$ is the Larmor frequency, the total lossless radio power (integrated over solid angle) emitted by electrons in a volume element $dV$ is \citep[e.g.][]{KDA+1997}
\begin{equation}
\begin{split}
dL_{\nu_{\rm syn}} = \frac{1}{2} \sigma_{\rm T} c u_{\rm B} \frac{\gamma^3}{\nu_{\rm syn}} n(\gamma) dV ,
\end{split}
\label{Lsyn}
\end{equation}
where $\sigma_{\rm T}$ is the electron scattering cross-section, $c$ is the speed of light, $u_{\rm B}$ is the energy density of the magnetic field, and $n(\gamma)$ is the electron energy distribution. The inverse-Compton power is similarly given by
\begin{equation}
\begin{split}
dL_\nu = \frac{1}{2} \sigma_{\rm T} c u_{\rm c} \frac{\gamma^3}{\nu} n(\gamma) dV ,
\end{split}
\label{Lic}
\end{equation}
where $u_{\rm c} = u_{\rm c0} (1 + z)^4$ is the energy density of the cosmic microwave background radiation at redshift $z$, $e$ the electron charge and $u_{\rm c0} = 0.25e\times10^6 \rm\, J \, m^{-3}$ is the corresponding energy density at the present epoch \citep{Longair+2010}. The electron population involved in both emission processes is identical, and thus the Lorentz factor and electron energy distribution can be modelled using the equations of \citet{Turner+2018a} derived for synchrotron radiation. For inverse-Compton emission, the relevant frequency is that of upscattered CMB photons,  $\nu = \gamma^2 \nu_{\rm cmb}$, where a photon at the peak CMB frequency $\nu_{\rm cmb} = 5.879\times10^{10} {\rm\, Hz \, K^{-1}} \times 2.73 {\rm\, K}(1 + z)$ is boosted by a synchrotron-emitting electron with a Lorentz factor $\gamma$.

The {peak} frequency of the synchrotron{-emitting electrons} involved in the inverse-Compton upscattering {of CMB photons to frequency $\nu$,} can thus be derived for rest-frame frequencies as
\begin{equation}
\begin{split}
\nu_{\rm syn} = \frac{3e \nu \sqrt{2 \mu_0 u_{\rm B}}}{2\pi m_{\rm e} \nu_{\rm cmb}} ,
\end{split}
\end{equation}
where $\mu_0$ is the permeability of free space and $m_{\rm e}$ the electron mass.

The full calculation of the inverse-Compton emissivity from the upscattering photons requires the integration over the total spectrum of CMB photons {(i.e. a blackbody distribution). The following correction should be applied to the simpler analytically tractable solution which assumes all CMB photons at a given redshift are at a single frequency \citep[e.g.][]{Nath+2010}:}
\begin{equation}
\begin{split}
J_{\rm corr}(s) = \frac{\pi^4}{15 \Gamma (\tfrac{s + 5}{2}) \zeta (\tfrac{s + 5}{2})} ,
\end{split}
\end{equation}
where $\Gamma$ and $\zeta$ are the Gamma and Zeta functions respectively. For a typical electron energy injection index of $s = 2.4$ (i.e. $N(\gamma) \propto \gamma^{-s}$), the {use of a single frequency CMB spectrum in this calculation would lead to an error of approximately 40\%}.

Following the method of \citet{Turner+2018a}, their Equation 3 is modified to yield the integrated inverse-Compton emissivity of the lobe at time $t$:
\begin{equation}
\begin{split}
L(\nu, t) = &\frac{K(s) {\nu_{\rm syn}}^{(1 - s)/2}}{J_{\rm corr}(s)} \left(\frac{\nu_{\rm syn}}{\nu} \right) \frac{q^{(s - 3)/4}}{(q + 1)^{(s + 1)/4}} \\
&\quad \times p(t)^{(s + 1)/4} V(t) (\Gamma_{\rm c} - 1) [u_{\rm c0} (z + 1)^4] \mathcal{Y}(\gamma, t),
\end{split}
\label{luminosityloss}
\end{equation}
where $K(s)$ is the source specific constant defined in Equation 5 of \citet{Turner+2018a}, $\Gamma_{\rm c}$ is the adiabatic index of the lobe plasma, $p(t)$ is the present-time lobe pressure, and $q \equiv u_B / u_e$ is the ratio of the energy density in the magnetic field to that in the synchrotron-emitting particles \citep{Turner+2018b}\footnote{RAiSE implicitly assumes a single magnetic field value throughout the lobes; although more complex configurations almost certainly exist \citep{Hardcastle2013, Hardcastle+2014} these are not well constrained at present, and our approach is standard for lobe modelling \citep[e.g.][]{Mocz+2011,Hardcastle+2018}.}. The differences between the synchrotron \citep[in][]{Turner+2018a} and inverse-Compton (here) forms of this equation can be explained as follows: (1) the power law spectrum is cast in terms of the synchrotron frequency rather than the (now inverse-Compton) emitting frequency, (2) the inverse-Compton frequency $\nu$ in the denominator of Equation \ref{Lic} replaces the synchrotron frequency in Equation \ref{Lsyn}, and (3) the inverse-Compton power in Equation \ref{Lic} replaces a linear dependence on the magnetic field energy density $u_{\rm B} = qp/(\Gamma_{\rm c} - 1)(q + 1)$ in Equation \ref{Lsyn} with a dependence on the energy density of the CMB radiation $u_{\rm c}$.

The loss function $\mathcal{Y}(\gamma, t)$ is defined in Equation 7 of \citet{Turner+2018} for active and remnant sources as
\begin{equation}
\begin{split}
\mathcal{Y}(t) = \int_0^t \frac{a_{\rm v}(t_{\rm i})}{t_{\rm i}}  \frac{Q(t_{\rm i})}{Q_0} \left[\frac{p(t_{\rm i})}{p(t)} \right]^{1 - 4/(3\Gamma_{\rm c)}} \frac{V(t_{\rm i})}{V(t)} \left[\frac{\gamma_{\rm i}}{\gamma} \right]^{2 - s} dt_{\rm i} ,
\end{split}
\label{resolvedvolume}
\end{equation}
where $Q(t_{\rm i})/Q_0$ is the instantaneous jet power at the electron injection time scaled by its active value \citep[i.e. 0 or 1 in this work;][]{Turner+2018}. The integral is over the particle injection times $t_{\rm i}$, where $p(t_{\rm i})$, $V(t_{\rm i})$ and $\gamma_{\rm i}$ are the lobe pressure, volume and electron Lorentz factor at the time of injection respectively. The constant $a_{\rm v}$ is the average volume expansion rate of the packet of electrons injected at time $t_{\rm i}$, defined through $V \propto t^{a_{\rm v}(t_{\rm i})}$.
The loss function is defined, as before, in terms of the Lorentz factor $\gamma$ of the synchrotron-emitting electrons which radiate at frequency $\nu_{\rm syn}$; the Lorentz factor of these electrons is the same as for the inverse-Compton upscattering process.

\begin{figure}
\begin{center}
\includegraphics[width=0.45\textwidth,trim={15 10 40 40},clip]{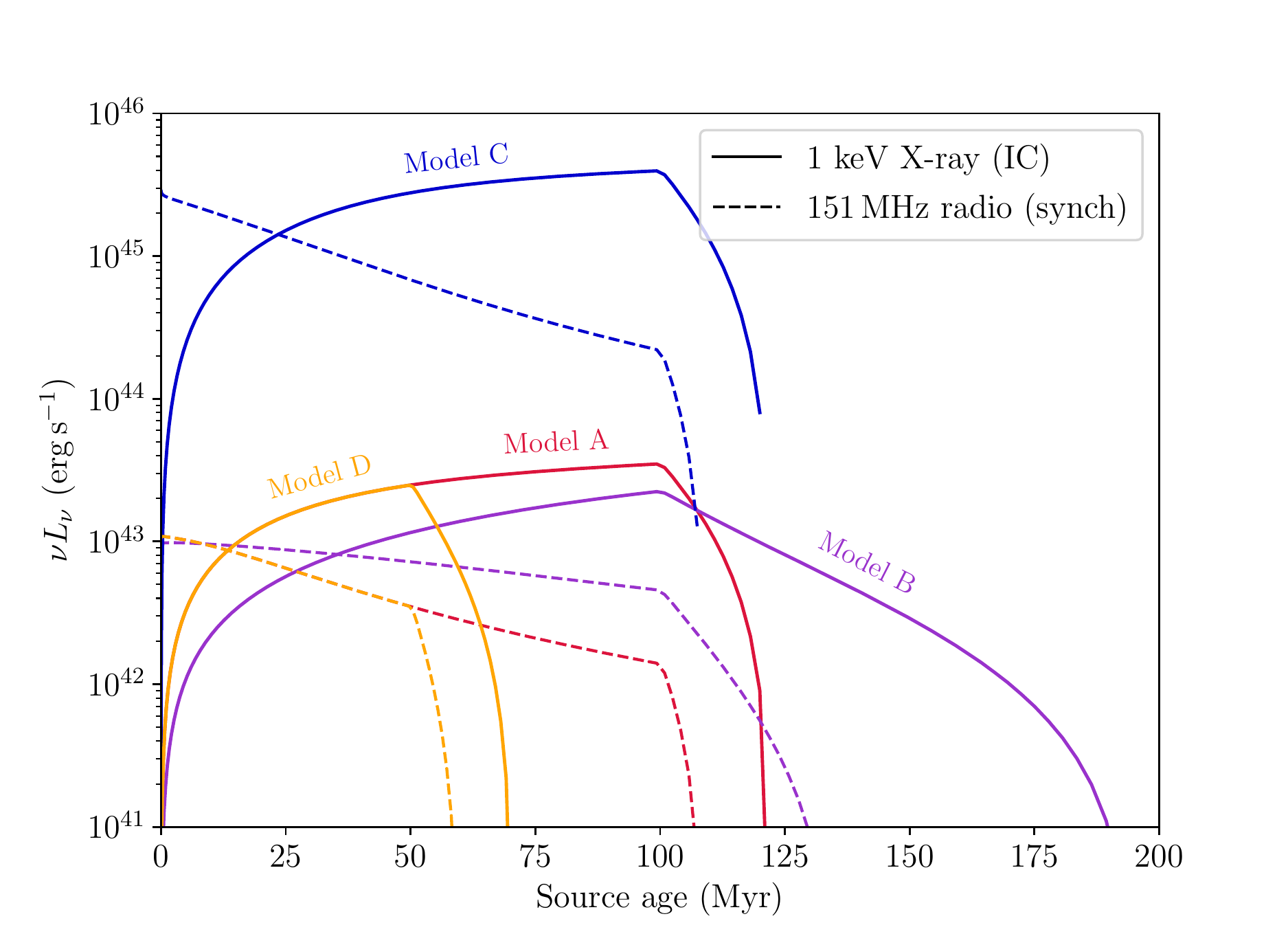}
\end{center}
\caption[]{{Luminosity--source age tracks for the evolution history of four extended sources at the observer-frame $151\rm\, MHz$ radio frequency (dashed lines) and $1\rm\, keV$ X-ray energy (solid lines). These luminosities are expressed as $\nu L_\nu$ for consistency with other authors. The four extended AGNs shown have the same jet powers, active lifetimes and redshifts as those considered in Figure 3 of \citet{Mocz+2011}. Model A (red): $Q = 3 \times 10^{38}\rm\, W$, $t = 100\rm\, Myrs$ and $z=2$; Model B (purple): same as Model A but $z=1$; Model C (blue): same as Model A but $Q = 3 \times 10^{40}\rm\, W$; Model D (orange): same as Model A but  $t = 50\rm\, Myrs$.}}
\label{fig:LDtracks}
\end{figure}

{The inverse-Compton emission is modelled over the evolutionary history of a source using the RAiSE dynamical model to derive the pressure and volume evolution of the lobe \citep{Turner+2015}. Luminosity--source age tracks are thus calculated at $1\rm\, keV$ observer-frame X-ray energies for four extended AGNs (Figure \ref{fig:LDtracks}). Our base model (Model A) considers a $Q = 3 \times 10^{38}\rm\, W$ jet with an active lifetime of $t = 100\rm\, Myrs$ at redshift $z = 2$. The gas density profile is modelled for a $10^{13.5}\rm\, M_\odot$ mass cluster following the method of \citet{Turner+2015}, the lobe axis ratio is set as $A = 4$, the injection index of the electron energy distribution is $s = 2.4$, whilst the other parameters in the RAiSE dynamical and inverse-Compton emissivity models take the same values as used by \citet{Turner+2018b}. Three variations to the base model are considered with either a lower redshift (Model B), higher jet power (Model C) or shorter active lifetime (Model D). The X-ray inverse-Compton luminosity--source age tracks for our four extended AGNs are qualitatively very similar (for example) to those in Figure 3 of \citet{Mocz+2011}, albeit with noticeably brighter luminosities, largely due to the richer cluster cores in our observationally informed density profiles.}

\subsection{Spatially resolved lobe losses}
\label{sec:SPATIALLY RESOLVED LOBE LOSSES}

The spatially resolved RAiSE dynamical model calculates lobe properties as a function of distance from the site of particle acceleration (hotspots for FR-II, flaring points for FR-I sources). Taking the particle acceleration site as the origin of the coordinate system, the inverse-Compton emissivity from a position $\textbf{r}$ in the lobe can similarly be derived following \citet{Turner+2018a}.
\begin{equation}
\begin{split}
dL(\nu&, t, \textbf{r}) = \frac{K(s) {\nu_{\rm syn}}^{(1 - s)/2}}{J_{\rm corr}(s)} \left(\frac{\nu_{\rm syn}}{\nu} \right) \frac{q^{(s + 1)/4}}{(q + 1)^{(s + 5)/4}} \\
&\times p(t, \textbf{r})^{(s + 1)/4} dV(t, \textbf{r}) (\Gamma_{\rm c} - 1) [u_{\rm c0} (z + 1)^4] \\
&\times \left[\frac{p(t_{\rm i}(\textbf{r}))}{p(t, \textbf{r})} \right]^{1 - 4/(3\Gamma_{\rm c})} \left[\frac{\gamma_{\rm i}(\textbf{r})}{\gamma} \right]^{2 - s}  ,
\end{split}
\label{resolvedluminosity}
\end{equation}
where the volume occupied at the present time by an electron packet injected between $t_{\rm i}$ and $t_{\rm i} + dt_{\rm i}$ is given by
\begin{equation}
\begin{split}
dV(t, \textbf{r}) &= \int_0^t \frac{Q(t_{\rm i})}{Q_0} \delta(t_{\rm i}' - t_{\rm i}) \frac{dV(t_{\rm i}', \textbf{r})}{dt_{\rm i}'} dt_{\rm i}' .
\end{split}
\label{resolvedvolume}
\end{equation}
For lobed sources with a high internal sound speed (typically FR-IIs), the volume element can be simplified to $dV(t, \textbf{r}) = a_{\rm v}(t_{\rm i}) V(t_{\rm i}) dt_{\rm i}/t_{\rm i}$ and the pressures taken as lobe averages; i.e. $p(t, \textbf{r})$ and $p(t_{\rm i}(\textbf{r}))$ are the lobe pressure at the present time and the time of electron injection respectively. 

\citet{Turner+2018a} used hydrodynamic simulations to study the spatial distribution of synchrotron-emitting electrons throughout lobed sources, by injecting tracer fields into the jet at regular intervals and tracking their motion. The distribution of these tracer fields within the lobe at each injection age was used to produce a representative synchrotron-emitting electron population at every location across the source.
\citet{Turner+2018a} found that the fluid injected at the jet terminal hotspot initially flows smoothly back towards the core, but electrons of different ages disperse broadly after travelling back slightly less than half of the lobe length. \citet{Turner+2018a} model the average location, $\textbf{r}$, and the 2$\sigma$ spread, $d\textbf{r}$, of the electron packets as a function of synchrotron age (see their Equations 11a,b). These authors also found the behaviour of the flow to be independent of both physical and temporal scales, enabling a single analytic description of the electron distribution to be assumed for all lobed FR-II sources.

Below, we use the results of these simulations to inform our analytic models for both active and remnant radio sources. For remnants, the relative locations of the electron packets within the lobe are assumed to remain fixed upon the jet switching off. This assumption is only valid on timescales shorter than the mixing timescales. However, remnant sources are typically observed soon after the jet switches off \citep[e.g. ][]{Brienza+2017,Hardcastle+2018,Turner+2018,Shabala+2019} and this assumption is expected to be appropriate for most objects.

\section{SHOCKED SHELL BREMSSTRAHLUNG RADIATION}
\label{sec:SHOCKED SHELL BREMSSTRAHLUNG RADIATION}

The bow shock generated by a powerful lobed radio source (FR-II or FR-I) sweeps up the intracluster medium (ICM) in its path as it propagates outwards from the active nucleus. The thermal evolution of this swept-up gas was considered by \citet{Alexander+2002}. Here, we extend the RAiSE model to explicitly include evolution of the shell of shocked gas surrounding the lobe.

\subsection{Shocked shell dynamical model}

\subsubsection{Axis ratio evolution}
\label{sec:AXIS_RATIO}

Hydrodynamical simulations \citep[e.g.][]{HK+2013} suggest {that} the bow shocks of radio sources expand in a self-similar manner despite the lobes slowly elongating over their evolutionary history: for example, in their Figure 1, the axis ratio of the shocked shell remains a constant value of $\sim 2.5$ once fully formed, whilst the lobe ratio slowly increases from 6.5 to 8. We therefore model the growth of the shocked shell in RAiSE using the same formalism as for lobe evolution, but excluding the late-stage Rayleigh-Taylor mixing which quickly pinches the lobes. The radius of the shocked shell is thus related to that of the intact lobe at each point on its surface as $R_{\rm s}(\theta) = \tau R(\theta)$, where $\tau \equiv \tau(\theta)$ is a constant of proportionality and $\theta$ is the angle between the surface location and the jet axis.

Simulations clearly show {that} the ratio between shocked shell and lobe radii varies across the surface of the shocked shell. We define the axis ratio (length divided by width) of the shocked shell in terms of the axis ratio of the lobe as $A_{\rm s} = A^\iota$, for some exponent $\iota$. Based on the hydrodynamical simulations of \citet{HK+2013}, we assume a value of $\iota = 0.5$ in this work; in other words, $A_{\rm s} = \sqrt{A}$.

The radial distance to the lobe surface at an angle $\theta$ from the jet axis is related to the length of the lobe along {the} jet axis by the ratio given in Equation 21 of \citet{Turner+2015},
\begin{equation}
\begin{split}
\eta(\theta) = \frac{1}{\sqrt{A^2 \sin^2 \theta + \cos^2 \theta}}
\end{split} ,
\label{eta}
\end{equation}
and the distance to the shocked shell at angle $\theta$ is similarly related to the length of the shocked shell along the jet axis by the ratio
\begin{equation}
\begin{split}
\eta_s(\theta) = \frac{1}{\sqrt{A^{2\iota} \sin^2 \theta + \cos^2 \theta}}
\end{split} .
\label{eta_s}
\end{equation}
The radius of the shocked shell can thus be related to that of the lobe as $R_s(\theta) = \tau \eta_{\rm s}(\theta) R(\theta)/\eta(\theta)$, where $\tau \sim 1.05$ \citep{HK+2013} is the ratio of the shocked shell to lobe radii along the jet axis.

\subsubsection{Modified shock geometry}

To describe the expansion of the shocked shell, we follow the geometric approach of \citet{Turner+2015}. We calculate the component of the expansion rate normal to the surface of the shocked shell by relating it to the expansion rate of the shocked shell along the jet axis, as in Equation 20 of \citet{Turner+2015},
\begin{equation}
\begin{split}
\zeta_s(\theta) = \left[\frac{A^{2\iota} \sin^2 \theta + \cos^2 \theta}{A^{4\iota} \sin^2 \theta + \cos^2 \theta} \right]^{1/2}
\end{split} .
\label{zeta_s}
\end{equation}
Again following the method of \citet{Turner+2015}, we derive a second order differential equation in terms of the lobe radius and expansion rate of the lobe surface at an angle $\theta$ from the jet axis. This differential equation cannot be solved analytically in general, and so we must adopt a numerical scheme using a fourth order Runge-Kutta method in terms of a system of two first order ODEs. The following system of equations must be solved for each small angular element $[\theta - d\theta/2, \theta + d\theta/2)$ of the lobe and shocked shell:
\begin{equation}
\begin{split}
\dot{R} &= v \\
\dot{v} &= \frac{3 (\Gamma_{\rm x} + 1)(\Gamma_{\rm c} - 1) Q R^{\beta - 3} d\lambda}{8 \pi v (\zeta_{\rm s}/\eta_{\rm s})^2 k \sin\theta d\theta} \left[\frac{\tau \eta_{\rm s}}{\eta} \right]^3 + \frac{(\beta - 3\Gamma_{\rm c}) v^2}{2 R} \\
&\quad\quad + \frac{(\Gamma_{\rm x} - 1) (3 \Gamma_{\rm c} - \beta - \xi) l}{4 R^{\xi + 1} (\zeta_{\rm s}/\eta_{\rm s})^2} ,
\end{split}
\label{supersonic system}
\end{equation}
where in the strong-shock supersonic limit (near time zero in the numerical scheme) the $\theta$ dependent constant $d\lambda$ is defined through the expression:
\begin{equation}
\begin{split}
&\frac{8\pi k \sin\theta d\theta}{3 (\Gamma_{\rm x} + 1)} \left[(3 \Gamma_{\rm c} - \beta)R^{2 - \beta} \dot{R}^3 + 2R^{3 - \beta} \dot{R} \ddot{R} \right]_{\theta = 0} \\
&\quad\quad \times {\eta_{\rm s}}^{3 - \beta}(\theta) {\zeta_{\rm s}}^2(\theta) \left[\frac{\eta(\theta)}{\tau \eta_{\rm s}(\theta)} \right]^3 = (\Gamma_{\rm c} - 1) Q\, d\lambda(\theta) .
\end{split}
\end{equation}
Here, $\beta$ and $k$ parametrise the {local} shape of the ambient density profile (i.e. $\rho = k r^{-\beta}$), $\xi$ and $l$ describe the {local} shape of the temperature profile (i.e. $T = l r^{-\xi}$), whilst $\Gamma_{\rm x}$ is the adiabatic index of the ambient medium. {The shape of density and temperature profiles are based on cluster observations and modelled as piecewise continuous power laws to retain analytically tractable solutions \citep{Turner+2015}.} The differential equations describing the evolution of the lobe in the subsonic and remnant phases are unchanged from those given in \citet{Turner+2015} and \citet{Turner+2018} respectively.

\subsection{Isothermal evolution of the shocked shell}

As first pointed out by \citet{Alexander+2002}, the final state of the shocked gas lying between the lobe surface and the bow shock depends critically on whether it expands adiabatically or isothermally; we assume here that the swept-up gas evolves isothermally. The pressure, $p_{\rm s}(\theta)$, of the swept-up gas at an angle $\theta$ from the jet axis is found from the jump conditions at the shock.

The cooling rate, and thus intensity of thermal bremsstrahlung radiation, in the shocked shell depends on both the pressure of the shocked gas and its temperature.
For low Mach numbers $M_0$ the cooling time of the swept-up gas, $t_{\rm s}$, can become short compared with that in the cluster, $t_{\rm x}$, with $t_{\rm s}/t_{\rm x} > 0.05M_0$ \citep{Alexander+2002}. However, the cooling time of the external gas would have to be at least an order of magnitude less than the source age, $t$, for the shocked gas to suffer significant radiative losses. Such conditions are only possible at the centres of the strongest cooling flow clusters \citep{Shabala+2009b} -- precisely \emph{not} the locations where powerful, large radio galaxies are typically found \citep{Miraghaei+2017}. We therefore ignore the cooling of the gas within the shocked shell in our analysis.

\subsubsection{Mean shocked shell temperature}

The mean temperature of the shocked gas is found by first calculating the mean pressure and density of the shocked shell. The mean density is taken as the ratio of the mass of ambient gas previously occupying the lobe and shocked shell to the volume of the shocked gas shell:
\begin{equation}
\begin{split}
\bar{\rho_{\rm s}}(t) = \frac{\int_0^{\pi/2} \int_0^{R_{\rm s}(t, \theta)} \rho(r) r^2 dr \sin \theta d\theta}{\int_0^{\pi/2} \int_{R(t, \theta)}^{R_{\rm s}(t, \theta)} r^2 dr \sin \theta d\theta} ,
\end{split}
\end{equation}
where $\rho(r)$ is the gas density of the ambient medium. This expression must be solved numerically except for the special case of spherical lobes expanding into a power law environment. 
{The mean temperature of the shocked gas shell is thus given as $T_{\rm s}(t) = \bar{m} \bar{p_{\rm s}}(t)/k_{\rm B} \bar{\rho_{\rm s}}(t)$ for a shocked gas shell with mean pressure $\bar{p_{\rm s}}(t)$ and an average particle mass $\bar{m} \sim 0.6m_{\rm p}$, where $m_{\rm p}$ is the proton mass. The temperature may of course vary throughout the shocked gas shell but without precise knowledge of any density gradients we simply assume mean values.}

\subsubsection{Bremsstrahlung radiation}

The X-ray emissivity per unit volume due to thermal bremsstrahlung radiation is \citep{Rybicki+1979}
\begin{equation}
\begin{split}
J(\nu) =& \frac{Z^2 e^6}{3 \pi^2 {\varepsilon_0}^3 {m_{\rm e}}^2 c^3} \left(\frac{\pi m_{\rm e}}{6} \right)^{1/2} \\
&\quad\quad (k_{\rm B} \mathcal{T})^{-5/2} {\tilde{p}}^{\!\:2} e^{-h\nu/k_{\rm B} \mathcal{T}} g(\nu, \mathcal{T}),
\end{split}
\end{equation}
where the pressure and temperature ($\tilde{p}$ and $\mathcal{T}$ respectively) may relate to the shocked shell, ambient medium, or any other plasma. Here, $Z \gtrsim 1$ is the average atomic number of the positively changed particles, $\varepsilon_0$ is the vacuum permittivity, and, at frequencies $h\nu \ll k_{\rm B} \mathcal{T}$, the Gaunt factor has a logarithmic dependence on frequency as
\begin{equation}
\begin{split}
g(\nu, \mathcal{T}) = \frac{\sqrt{3}}{\pi} \ln \left(\frac{4}{\zeta} \frac{k_{\rm B} \mathcal{T}}{h\nu} \right) ,
\end{split}
\end{equation}
where here $\zeta = 1.78$ is Gauss' number. In this work, we assume the metallicity of the plasma is 0.3 times the solar value, corresponding to an average atomic number of $Z \sim 1.04$. The shocked gas pressure used in the bremsstrahlung radiation calculation is that derived from the shock jump conditions on the shell surface at angle $\theta$ (i.e. $\tilde{p} = p_{\rm s}(\theta)$), whilst the mean temperature is derived following the method described in the previous section.

\begin{figure}
\begin{center}
\includegraphics[width=0.45\textwidth,trim={20 15 45 45},clip]{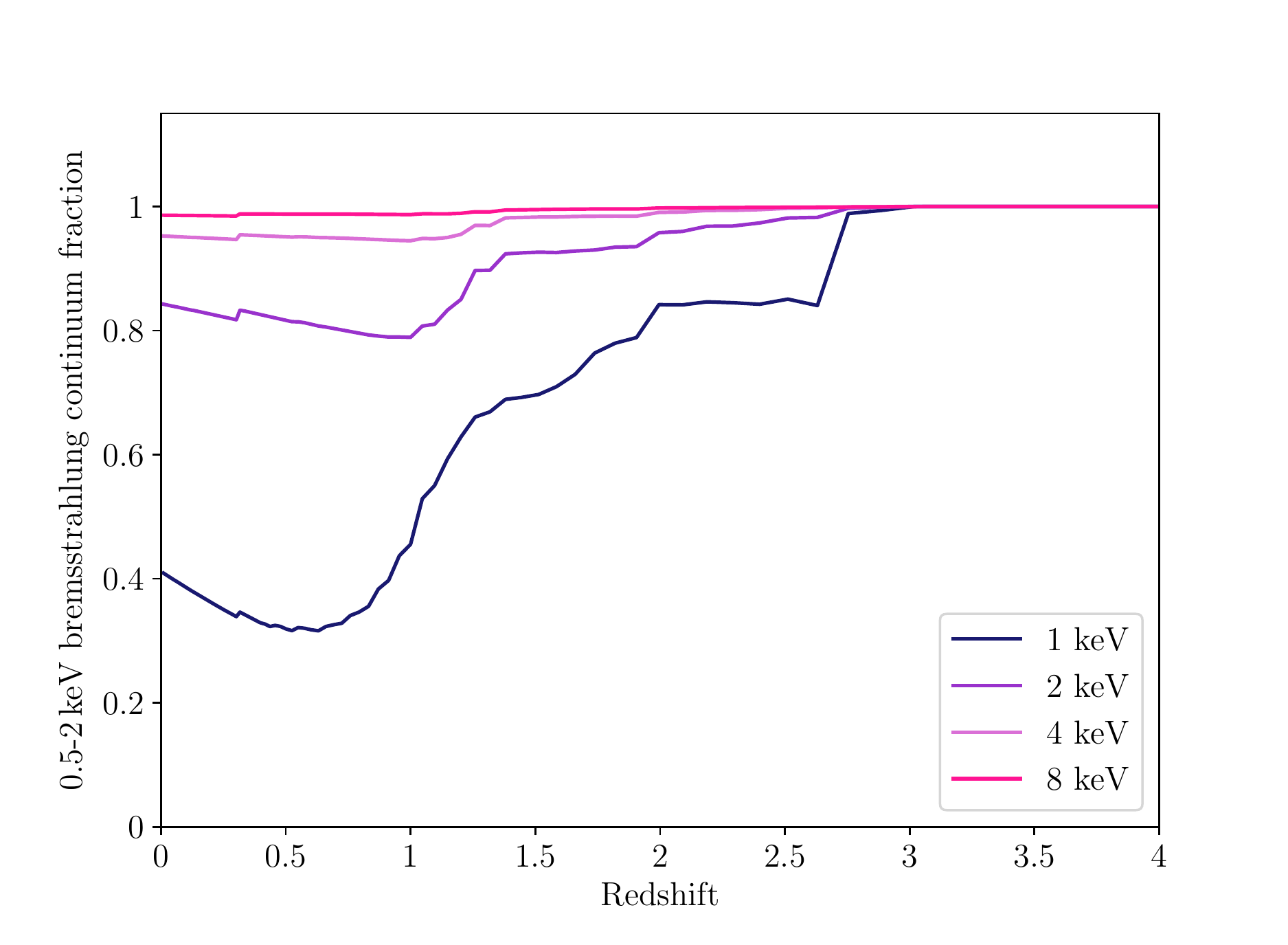}
\end{center}
\caption[]{{Fractional contribution of continuum emission to the total bremsstrahlung X-ray spectrum (continuum and emission lines) integrated over the 0.5-$2\rm\, keV$ band. The emission line spectra are calculated using the APEC thin-thermal plasma model as a function of redshift for $kT = 1$, 2, 4 and $8\rm\, keV$, assuming a metallicity of 0.3 times the solar value.}}
\label{fig:apec}
\end{figure}

{The X-ray spectra of smaller clusters at low-redshift are dominated by emission lines at the wavelengths of the \emph{Chandra} and \emph{eRosita} surveys. The APEC plasma code \citep{Smith+2001} can calculate the X-ray emission spectra from thin-thermal plasma for various temperatures and metallicities. In Figure \ref{fig:apec}, we plot the fractional contribution of continuum emission to the total bremsstrahlung X-ray spectrum (including both continuum and line emission) integrated over a typical 0.5-$2\rm\, keV$ observing band. The emission lines are found to contribute a significant component of the X-ray flux density at redshifts $z < 2.5$ in clusters with temperatures $kT \lesssim 1\rm\, keV$, corresponding to halo masses $\lesssim 10^{13.5}\rm\, M_\odot$. We use the appropriate APEC model to correct our analytically derived flux densities of the X-ray emission arising from bremsstrahlung radiation.}

\section{X-RAY SURFACE BRIGHTNESS MAPS}
\label{sec:X-ray surface brightness maps}

We use the formalism developed in the preceding sections to make predictions for the observed X-ray emission from radio galaxies and their environments. There are three main contributions: lobe inverse-Compton emission (Section \ref{sec:LUMINOSITY MODEL}) and two sources of bremsstrahlung radiation, from the shocked gas shell ahead of the lobes (Section \ref{sec:SHOCKED SHELL BREMSSTRAHLUNG RADIATION}), and from the ambient medium. We calculate the latter in the same way as bremsstrahlung from the shocked gas shell, but adopting density and temperature parameters characteristic of the hot gas in galaxy clusters, as described in \citet{Turner+2015}.

\subsection{X-ray and radio brightness maps}
\label{sec:Surface brightness}

The lobes and shocked gas shell of each mock radio galaxy are divided into a $512\times512\times512$ grid of cubic pixels; this grid extends a factor of two beyond the edge of the lobe to enable a comparison with X-ray surface brightness of the ambient medium. Each cell in the cube is classified as either part of the lobe, shocked shell or ambient medium based on the modified RAiSE dynamical model (Section \ref{sec:SHOCKED SHELL BREMSSTRAHLUNG RADIATION}), and the inverse-Compton emissivity or bremsstrahlung radiation calculated as appropriate. The observed emission arising from these extended objects is the integral of all the emissivity along a given line-of-sight. 
The two-dimensional surface brightness is simply calculated by summing the emissivity from every cell along the depth of the source, assuming the lobe plasma and ambient medium in front of the source is optically thin. The X-ray emission arising from the ambient medium lying outside the grid of cell is calculated analytically for each line-of-sight and added to the total from the numerical grid. For simplicity, we assume that the lobes lie in the plane of the sky.

The inverse-Compton and bremsstrahlung emissivity in the (two-dimensional) surface brightness map are derived for the technical characteristics of the \textit{extended Roentgen Survey with an Imaging Telescope Array} \citep[eRosita;][]{Merloni+2012}. Specifically, at a $1\rm\, keV$ {observer-frame energy}, the half-energy width (on axis) is $15\rm\, arcsec$ and the total effective area of the seven mirror systems is $\sim 1500\rm\, cm^2$. The number of $1\rm\, keV$ photons falling on these mirrors in a typical 1000$\rm\, s$ (1 ks) observing time is thus calculated from the X-ray surface brightness grid. The $151\rm\, MHz$ radio frequency emission from the lobes is also calculated following the method of \citet{Turner+2018a}; this frequency is commonly used by low-frequency Square Kilometre Pathfinder instruments, including the \textit{Low-Frequency Array} (LOFAR) and the \textit{Murchison Widefield Array} (MWA).

The X-ray and radio surface brightness distributions are modelled for lobed FR-IIs at a range of redshifts throughout their evolutionary history. Specifically, we simulate AGNs at 41 (log-spaced) source ages between 3 and 300$\rm\, Myrs$, redshifts of $z = 0.1$, 0.5 and 1, jet powers of $Q = 10^{37}$, $10^{38}$ and $10^{39}\rm\, W$, active ages of 30 and 100$\rm\, Myrs$ (i.e. when the jet ceases injecting fresh electrons), and host cluster environments with dark halo masses of $10^{12.5}$, $10^{13.5}$ and $10^{14.5}\rm\, M_\odot$. The axis ratio of the lobe is taken as $A = 4$ corresponding to an axis ratio for the shocked shell of $A_{\rm s} = 2$ (see Section~\ref{sec:AXIS_RATIO}). The other parameters in the RAiSE dynamical and synchrotron/inverse-Compton emissivity models take the same values as used by \citet{Turner+2018b}. 

\begin{figure}
\begin{center}
\includegraphics[width=0.45\textwidth,trim={20 10 0 40},clip]{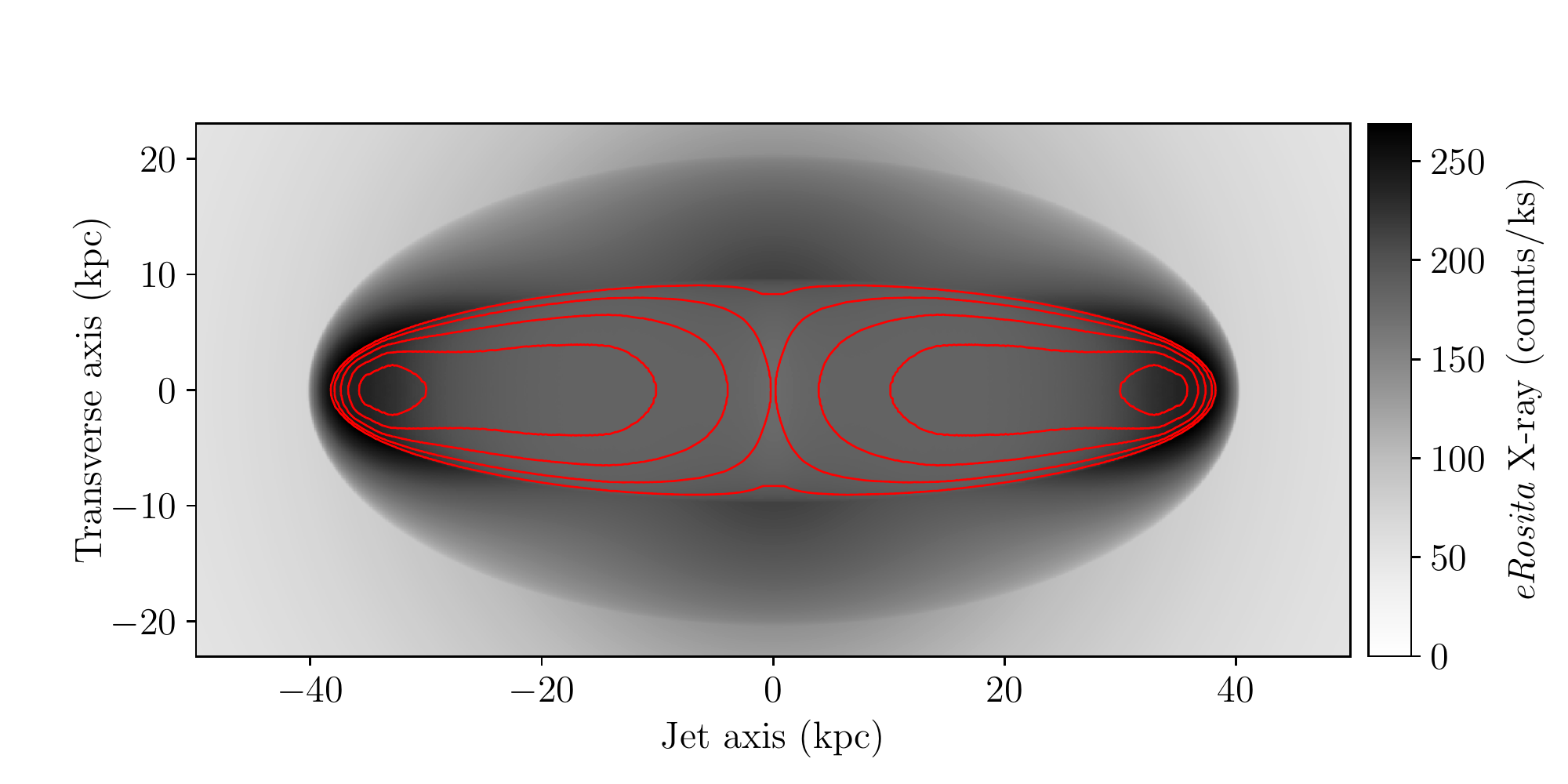} \\ \vspace{0.1cm}
 \includegraphics[width=0.45\textwidth,trim={20 10 0 40},clip]{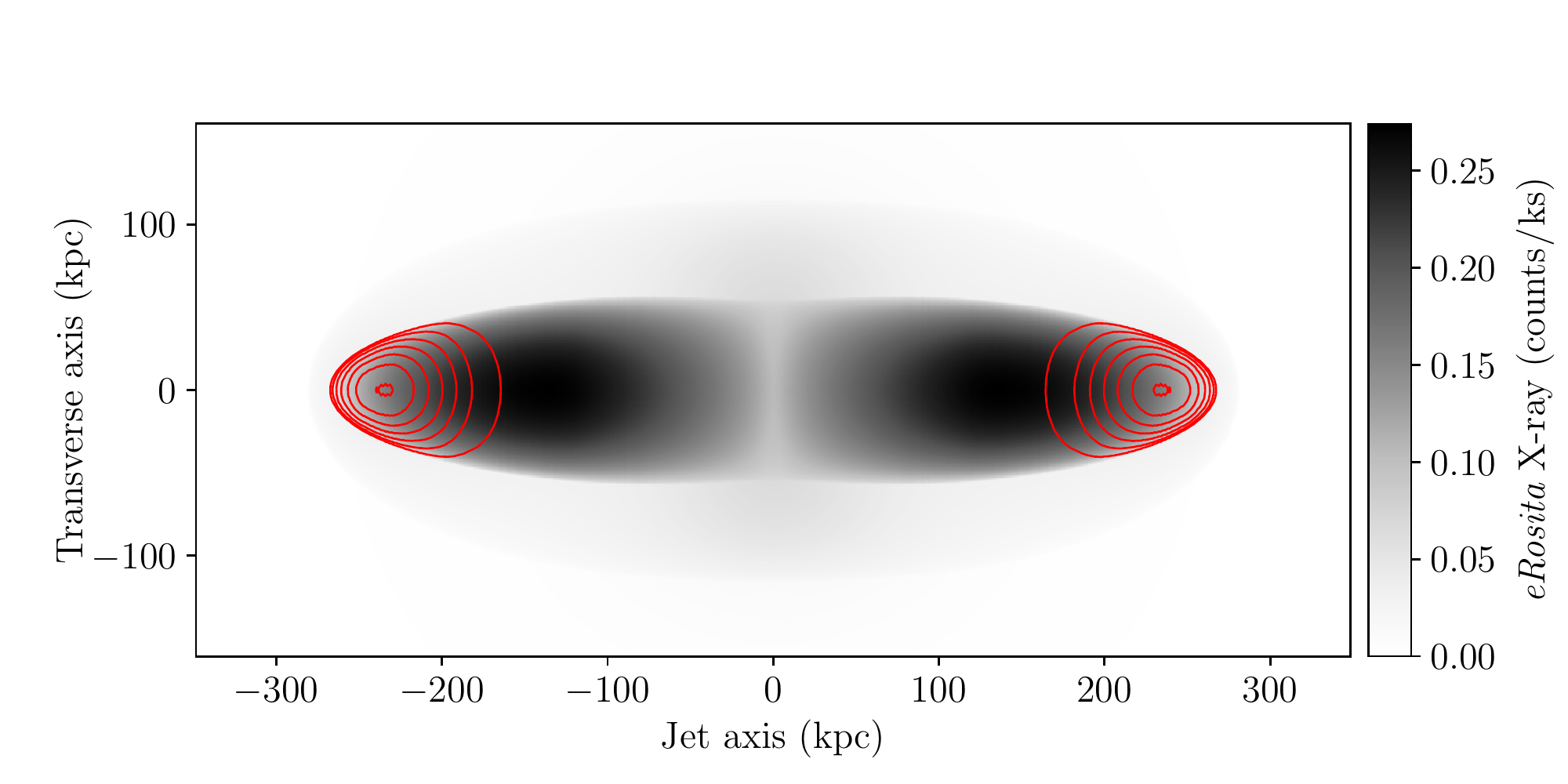} \\ \vspace{0.1cm}
\includegraphics[width=0.45\textwidth,trim={20 10 0 40},clip]{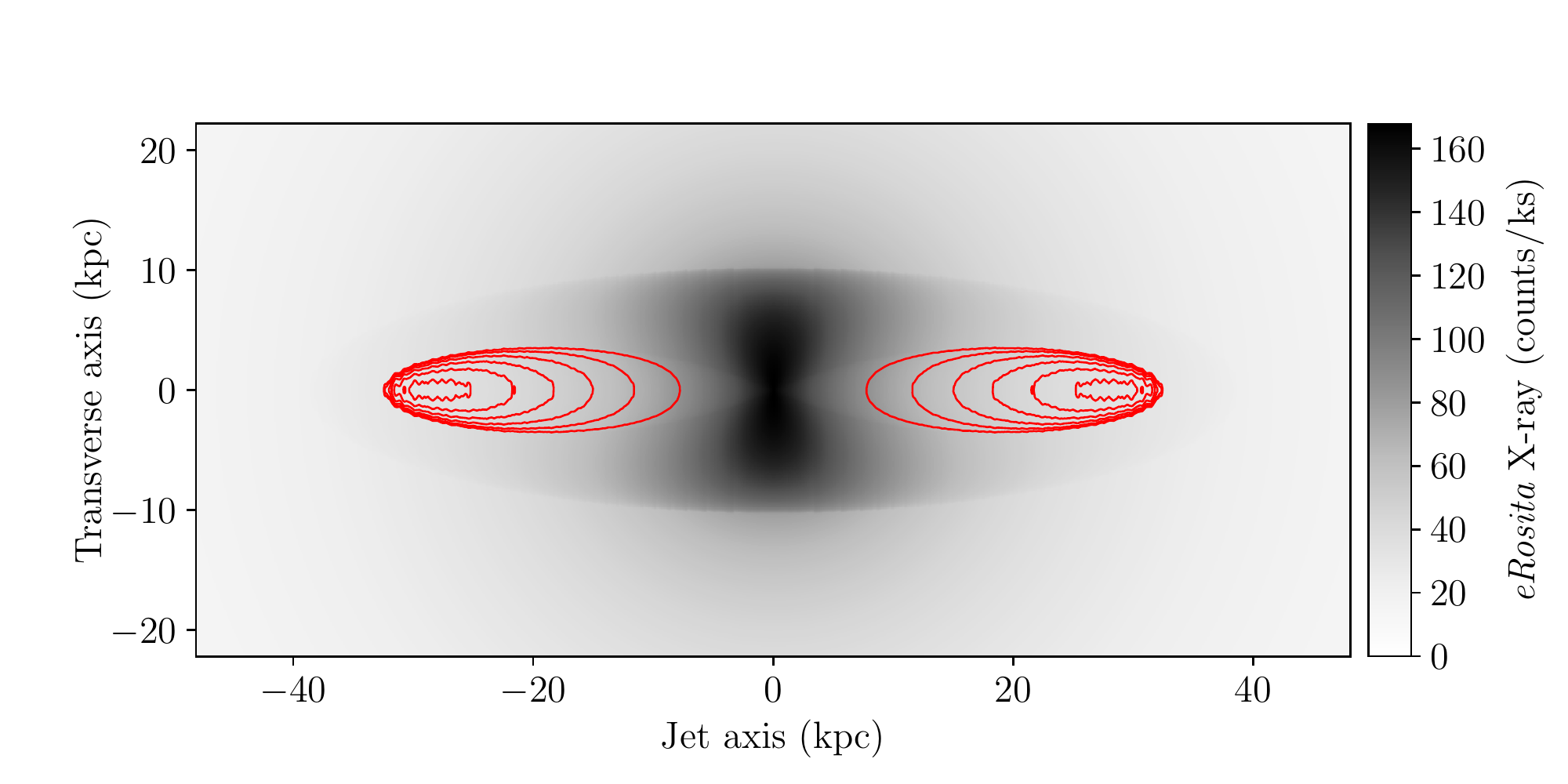} 
\end{center}
\caption[]{X-ray (1$\rm\, keV$) and radio (151$\rm\, MHz$) {observer-frame} frequency surface brightness maps for three simulated lobed AGNs. The X-ray surface brightness {(in greyscale)} is shown for the 15$\rm\, arcsec$ beam of eRosita, though the original resolution of the simulated image is retained for clarity. The radio frequency contours {(in red)} are linearly spaced between zero and the peak flux density; the radio contours are only shown here to highlight the sections of the lobe emitting synchrotron radiation and are thus not scaled for any particular survey. Top panel: active source with a $10^{39}\rm\, W$ jet expanding into a $10^{14.5}\rm\, M_\odot$ cluster at redshift $z = 0.1$, viewed at an age of $10\rm\, Myrs$. Middle panel: remnant source which had a powerful $10^{38}\rm\, W$ jet expanding into a $10^{12.5}\rm\, M_\odot$ cluster at redshift $z = 1$ for $100\rm\, Myrs$, viewed at an age of $125\rm\, Myrs$. Bottom panel: active source with a $10^{37}\rm\, W$ jet expanding into a $10^{14.5}\rm\, M_\odot$ cluster at redshift $z = 1$, viewed at an age of $100\rm\, Myrs$.}
\label{fig:surfbright}
\end{figure}

The surface brightness maps for several informative combinations of these parameters are shown in Figure \ref{fig:surfbright}. Powerful jets expanding into dense environments are found to create dense shells of shocked gas surrounding the lobe. The bremsstrahlung radiation from the shocked gas is much brighter than can be generated by the inverse-Compton radiation process, especially in young sources and at low redshifts (e.g. top panel of Figure \ref{fig:surfbright}). The faint lobes bounded by well-defined lines are seen in \textit{Chandra} X-ray images of Cygnus A and MS\,0735.6+7421 \citep{Rafferty+2006}, with cavities also seen in the hosts of weaker FR-Is such as the Perseus cluster \citep{Fabian+2006}. By contrast, in lower mass clusters, the density of swept-up gas is very low causing its temperature to be high to satisfy the shock jump conditions. The increased temperature leads to minimal bremsstrahlung radiation from the shocked gas shell. The inverse-Compton upscattered photons in the lobe are thus the dominant source of X-ray emission, especially at higher redshifts where the cosmic microwave background radiation is stronger (e.g. middle panel of Figure \ref{fig:surfbright}). Finally, the X-ray inverse-Compton emission remains visible well after the radio frequency emission vanishes; in the simulated remnant, the radio emission has retreated towards the freshest electrons at the hotspot, whilst the X-ray emission does a much better job of tracing out the channel evacuated by the radio lobes. A similar result has previously been reported by \citet{Mocz+2011}; Figure~\ref{fig:surfbright} additionally shows that the peak in the X-ray intensity from the lobe is much closer to the core than the full extent of the radio source. Many radio sources appear asymmetric, due to asymmetry in the environments encountered by the two lobes \citep{Rodman+2019}; this poses a challenge for robustly identifying host galaxies of remnant radio sources. Our results suggest that X-ray observations may provide a useful alternative.

An intriguing surface brightness map is produced when propagating weak jets into dense environments (bottom panel of Figure \ref{fig:surfbright}); dense shells of shocked gas build up around the lobe as before, however, the weakened shocks and magnetic fields allow the contact surface near the core to become unstable to turbulent mixing of the dense shell and the lobe. This results in thermal bremsstrahlung emission extending out from the core along the transverse axis, brighter than X-ray emission from the lobe or the thinner (and thus fainter integrated along the line-of-sight), unmixed portions of the shocked shell further from the core. We note that strong X-ray emission may also be observed {\it along} the jet axis due to inverse-Compton emission from the jets as well as lobes (for older sources at high redshift); a combination of radio and X-ray imaging may distinguish these two scenarios.

\subsection{Dominant source of X-ray emissivity}
\label{sec:Source of X-ray emissivity}

In this section, we investigate the relative importance of contributions to the X-ray surface brightness from the lobe (inverse-Compton), shocked gas shell (bremsstrahlung), and the ambient medium (bremsstrahlung), for some representative sources at a range of redshifts. The surface brightness for the three regions is calculated as the maximum value along the jet axis. However, the increased bremsstrahlung radiation from sources pinched by Rayleigh-Taylor mixing may present a source of error in this work as the rate of mixing is a poorly quantified model parameter in RAiSE \citep{Turner+2015}. Fortunately, in the vast majority of extended AGNs this enhanced emission is only a minor contributor to the integrated X-ray luminosity, and when generating surface brightness maps, either the Rayleigh-Taylor mixed region is not resolved from the bright X-ray core (discussed in Section \ref{sec:X-ray source number densities}) or the remainder of the lobe also sits above the surface brightness sensitivity limit (i.e. the shape of the source is correctly determined). The analyses in this work are repeated with the inner third of the lobe masked to exclude this region of enhanced emission; our results are unaffected by this change.

\begin{figure}
\begin{center}
\includegraphics[width=0.45\textwidth,trim={10 50 40 80},clip]{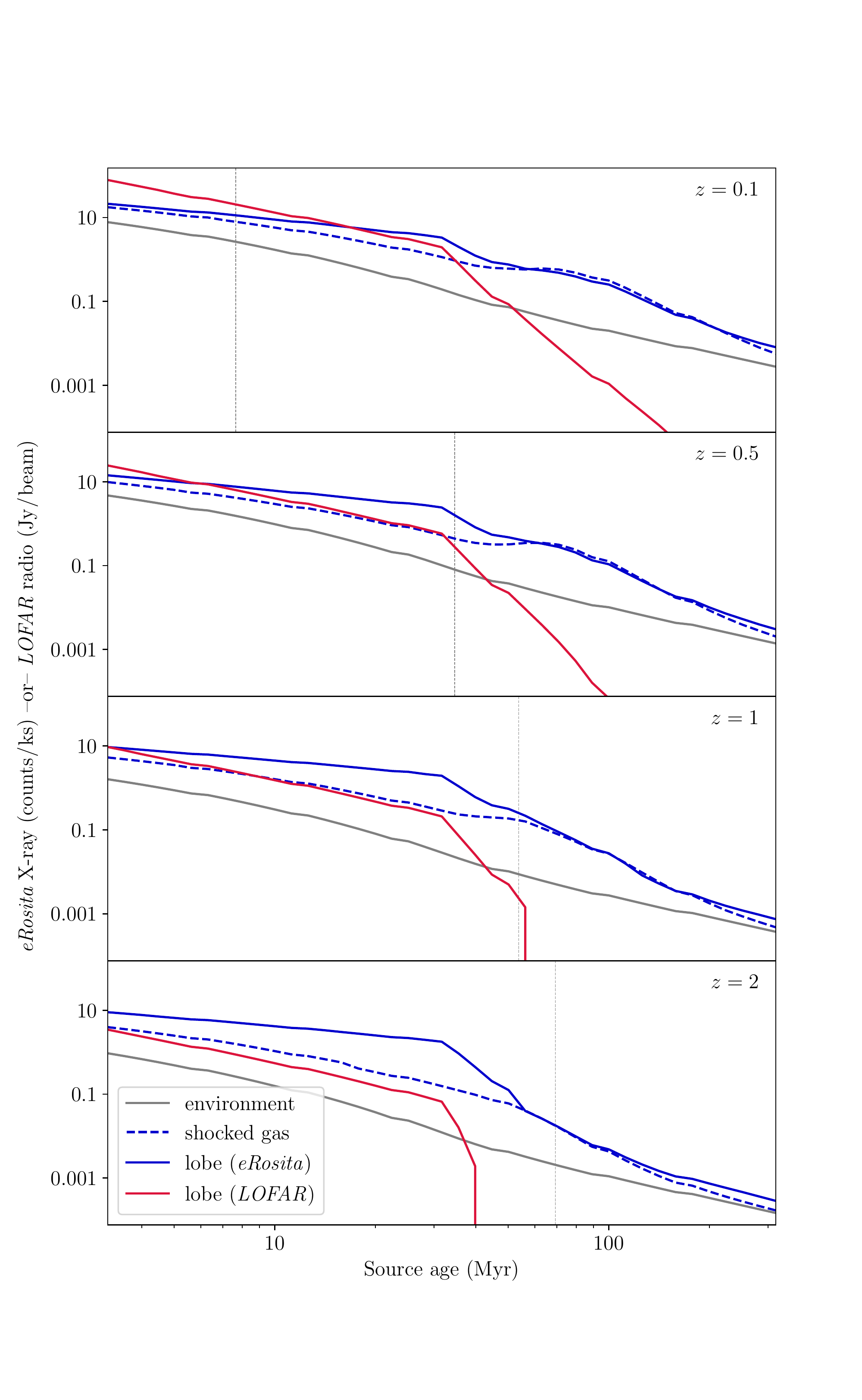}
\end{center}
\caption[]{Maximum X-ray (1$\rm\, keV$ {observer-frame}) surface brightness along the jet axis in the lobe (solid blue -- inverse-Compton), shocked gas shell (dashed blue -- bremsstrahlung) and the ambient medium (solid grey -- bremsstrahlung). The maximum radio (151$\rm\, MHz$ {observer-frame}) surface brightness in the lobe is also plotted for comparison (solid red -- synchrotron). The tracks show the surface brightness evolution for a remnant source powered by a $Q = 10^{38}\rm\, W$ jet that is active for $30\rm\, Myrs$, expanding into a $10^{12.5}\rm\, M_\odot$ halo mass cluster, at redshifts $z = 0.1$, 0.5, 1 and 2. The $\sim$$15\rm\, arcsec$ \emph{eRosita} resolution is shown for comparison using dashed vertical lines.}
\label{fig:timeseries1}
\end{figure}

\begin{figure}
\begin{center}
\includegraphics[width=0.45\textwidth,trim={10 50 40 80},clip]{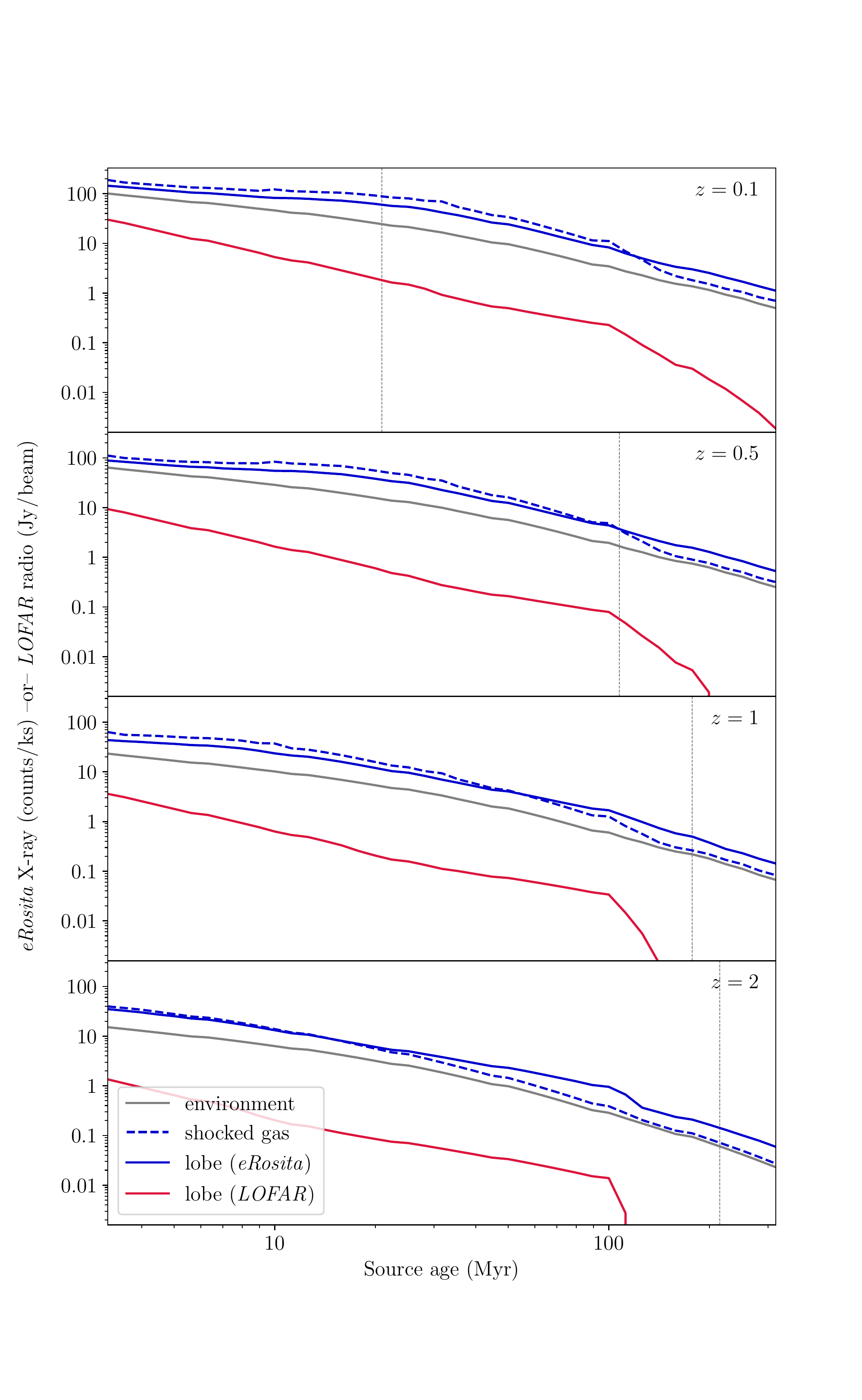}
\end{center}
\caption[]{Same as Figure \ref{fig:timeseries1}, but for a moderate $Q = 10^{37}\rm\, W$ kinetic power jet, active for $100\rm\, Myrs$, and expanding into a $10^{13.5}\rm\, M_\odot$ halo mass cluster.}
\label{fig:timeseries2}
\end{figure}

The evolutionary tracks of the X-ray surface brightness for these representative sources are shown in Figures \ref{fig:timeseries1} and \ref{fig:timeseries2}. The inverse-Compton emission is dominant over the bremsstrahlung radiation from both the shocked gas shell and the ambient medium in poor clusters ($10^{12.5}\rm\, M_\odot$; Figure~\ref{fig:timeseries1}) at all source ages $\gtrsim 1\rm\, Myr$. By contrast, the shocked gas shell of sources expanding into denser host cluster environments ($\gtrsim 10^{13.5}\rm\, M_\odot$; Figure~\ref{fig:timeseries2}) is brighter than the lobe, albeit not appreciably, for at least 20 to $100\rm\, Myrs$ with the weakest FR-II jet powers ($Q = 10^{37}\rm\, W$); the shocked shell stays brighter than the lobe for longer with either denser environments or higher jet powers. This change in the importance of the lobe and shocked gas shell to the X-ray surface brightness typically results from rapidly falling pressure in the shocked gas as the source approaches pressure equilibrium with the ambient medium. Further, the stronger cosmic microwave background radiation at higher redshifts leads to increased inverse-Compton emission from the lobe causing this source of X-ray surface brightness to become dominant in younger sources.

Finally, the X-ray surface brightness from the lobe and shocked gas shell remain detectable above that of the ambient medium for a considerable time after the jet switches off. {The extended source in Figure \ref{fig:timeseries1}, with an active lifetime of 30$\rm\, Myrs$, remains detectable over the ambient medium (at the $5\sigma$ level) for a further 155$\rm\, Myrs$ at redshift $z=0.1$, 115$\rm\, Myrs$ at $z=0.5$, 100$\rm\, Myrs$ at $z=1$, and 65$\rm\, Myrs$ at $z=2$. By contrast, \emph{any} level of synchrotron emission is present at radio frequencies for only 170, 70, 25 and 10$\rm\, Myrs$ after the jets switch off, respectively, at these redshifts. Similarly, the extended source in Figure \ref{fig:timeseries2} emits radiation from the lobe and shocked shell at X-ray wavelengths long after the radio emission fades, though this is only a factor of a couple brighter than the ambient medium.} 
{Based on these and other test cases,} we expect a population of remnant extended radio galaxies undetectable at radio frequencies but visible to X-ray telescopes, in line with literature predictions of a large population of inverse-Compton `ghosts', particularly at high-redshift \citep{BR+1999, Mocz+2011}.

\section{MOCK EXTENDED AGN POPULATION}
\label{sec:MOCK EXTENDED AGN POPULATION}

We now extend our analysis to make predictions for radio galaxy population statistics. Specifically, we combine the theoretical framework developed in the previous section with the halo mass function and literature radio frequency observations to create a mock population of extended radio galaxies. We then use this simulated population to predict the X-ray luminosity function for extended AGN emission, and investigate the number of objects that could be uniquely detected using large sky X-ray surveys.

\subsection{Construction of mock catalogue}

We construct a mock catalogue of extended AGNs, including their angular sizes, surface brightnesses, and integrated flux densities, by running simulations for a dense grid of model parameters. The brightnesses are calculated at radio frequencies for a typical \emph{Low Frequency Array} (LOFAR) survey and in the X-ray for an \emph{eRosita} survey. Observations are used to constrain the input halo mass function for AGN hosts, and the distribution of source active lifetimes and jet kinetic powers. For the remaining parameters, we assume values given in Section \ref{sec:Surface brightness}.

\subsubsection{Halo mass function}

\begin{figure*}
\begin{center}
\includegraphics[width=0.90\textwidth,trim={85 5 100 35},clip]{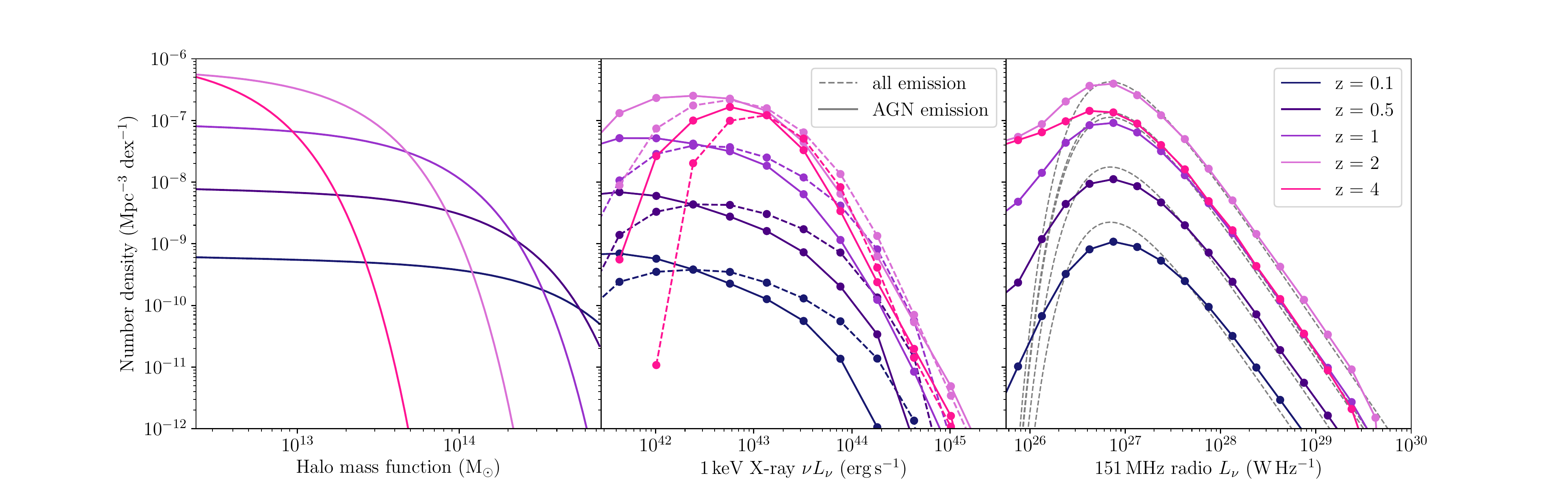}\\
\end{center}
\caption[]{Luminosity functions simulated for the mock population of extended AGNs in their active phase. Left: the halo mass function taken from \citet{Girardi+2000}, and extended to higher redshifts using SAGE \citep{Croton+2016}, is convolved with {a theoretical} AGN duty cycle \citep[e.g.][]{Pope+2012} to obtain the mass function for hosts of AGNs with FR-II morphology (solid lines). The halo mass function is shown at redshifts $z = 0.1$, 0.5, 1, 2 and 4. Centre: the \emph{eRosita} $1\rm\, keV$ {observer-frame} X-ray luminosity function derived from the mock population based on the halo mass function, and jet power and active age distributions discussed in the text. The luminosity function is shown including either only the emission associated with the AGN outburst (i.e. lobe and shocked shell; solid lines), or emission from both the extended AGN and ambient medium (dashed lines). Right: the \emph{LOFAR} $151\rm\, MHz$ {observer-frame} radio lumninosity is similarly calculated from the mock AGN population at the four redshifts. The dashed black lines are the observed $151\rm\, MHz$ radio luminosity function for powerful FR-IIs in the 3CRR, 6CE and 7CRS samples \citep{Willott+2001}.}
\label{fig:lumin_functions}
\end{figure*}

\begin{figure}
\begin{center}
\includegraphics[width=0.45\textwidth,trim={10 50 40 80},clip]{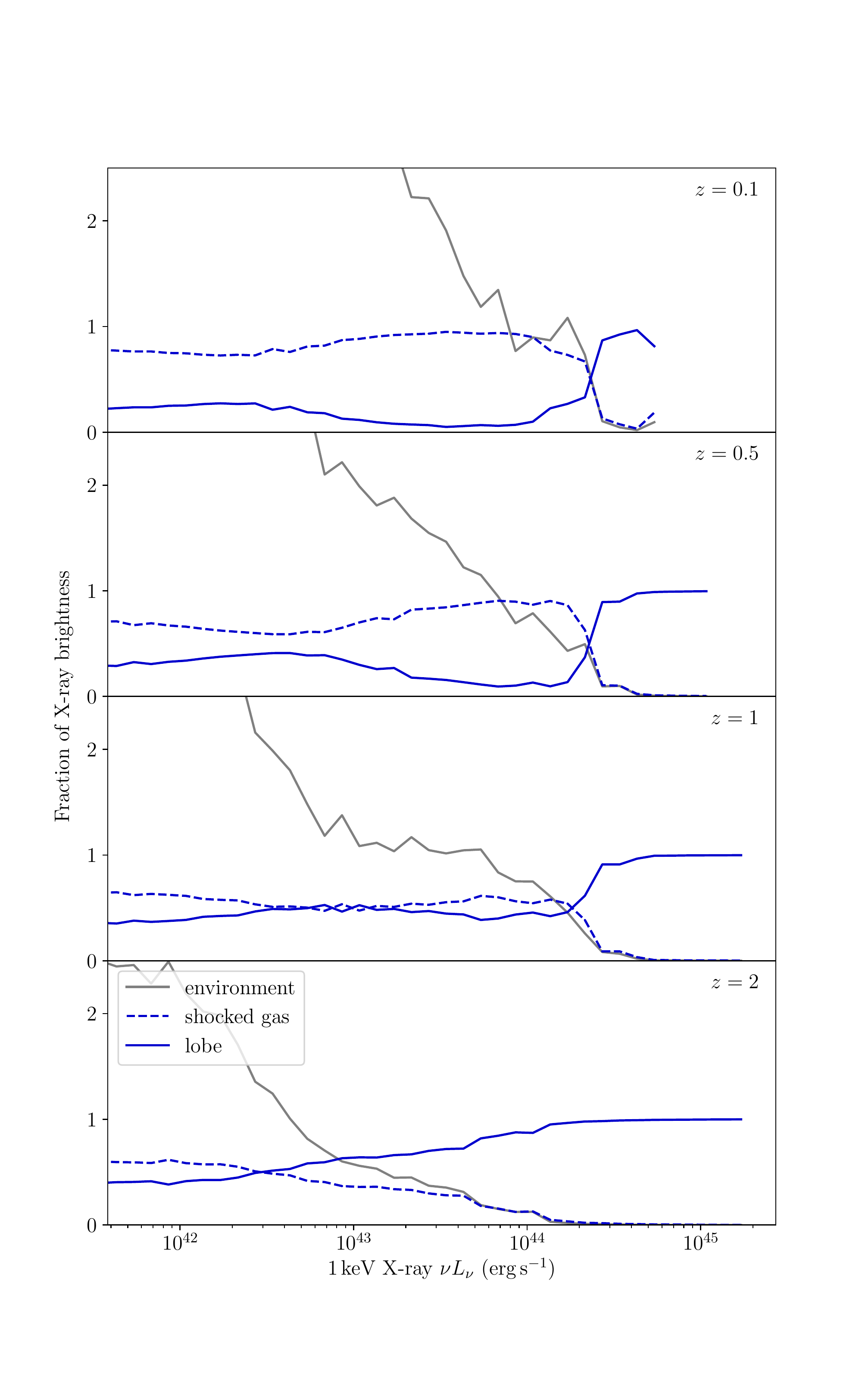}
\end{center}
\caption[]{Fractional contribution of the different radiative mechanisms to the total integrated X-ray luminosity, shown as both a function of luminosity and redshift. The inverse-Compton emission from the lobe is shown by the solid blue line, the bremsstrahlung radiation from the shocked gas shell surrounding the lobe by the dashed blue line, and the bremsstrahlung emission from the ambient medium is plotted in grey. The vertical axis is scaled so the emission from the extended AGN (i.e. lobe and shocked shell) sums to unity.}
\label{fig:lumin_fractions}
\end{figure}

The mass function of extended AGN hosting clusters is assumed to be the convolution of the halo mass function for all groups and clusters, and the probability of finding an AGN in a given mass host. The mass of dark matter haloes (observed as galaxy groups and clusters) can in general be described by a mass function, which gives the number density of clusters as a function of mass \citep[e.g.][]{Reiprich+2002}. In this work, and following \citet{Hardcastle+2019}, we take our dark matter halo masses from the low-redshift mass function of \citet{Girardi+2000} who find that a common Schechter function describes both galaxy groups and clusters. We use the semi-analytic galaxy evolution (SAGE) model of \citet{Croton+2016} to extend their observations to higher redshifts, finding the mass of the break in the Schechter function scales with redshift as approximately $(1 + z)^{-3}$. 

The radio-loud fraction of AGNs (i.e. fraction of sources with brightness above an arbitrary cut in radio luminosity) suggests hosts with higher mass black holes will have more frequent (or longer) phases of activity \citep[e.g.][]{Best+2005,Sabater+2019}. \citet{Pope+2012} made a theoretical prediction for the AGN duty cycle (i.e. the fraction of time a given source is active) as a function {of} black hole mass, $\delta \propto \rm M_\bullet^{1.5}$; {this relationship is consistent with the observed radio-loud fraction for all but the most massive galaxies in which the duty cycle peaks at 100\% \citep{Sabater+2019}}. The black hole mass in the brightest cluster galaxy (BCG), the mass of the stellar bulge, and the dark matter halo mass, are all known to scale approximately linearly with each other \citep{Magorrian+1998, Haring+2004, Gultekin+2009}. {Individual galaxies show moderate scatter of $0.5\rm\, dex$ about the dark matter halo--black hole mass relationship \citep[derived from][]{Haring+2004}, however the underlying relationship remains when considering several orders of magnitude in halo mass.} We therefore convolve the halo mass function with the mass--duty cycle relation to derive the radio AGN mass function. This halo mass function is shown in the left panel of Figure \ref{fig:lumin_functions} as a function of redshift for masses between $10^{12.5}$ and $10^{14.75}\rm\, M_\odot$. The number density is scaled in the plot based on the number of high-luminosity radio AGNs of FR-II morphology observed by \citet[][see Section \ref{sec:Jet kinetic powers}]{Willott+2001}.

These scaling relationships predict that all plausible host halo masses occur at similar probability, with a slightly decreased likelihood of finding a radio AGN in the most massive clusters. This convolved mass function will provide a better description of the environments typically hosting AGNs than the raw halo mass function.

\subsubsection{Source lifetimes}

The radio AGN lifetime function has generated much recent interest. The large observed fractions of compact sources \citep{Shabala+2008, Hardcastle+2019} strongly suggests a dominant population of short-lived radio jets. Recently, \citet{Hardcastle+2019} used a combination of data from the LOFAR LoTSS survey and dynamical models, to show that the majority of radio AGNs are consistent with models in which the source lifetime distribution is log-uniform. \citet{Shabala+2019} applied self-consistent modelling to LOFAR observations of active, remnant and restarted radio sources in the HETDEX field to similarly infer a dominant short-lived population, consistent with feedback-regulated accretion \citep{Novak+2011, Gaspari+2017}.

In line with these results, we similarly adopt here a log-uniform distribution of the active lifetimes between 3 and 300\,Myrs, and assume that the duty cycle is (on average) independent of the active lifetime. We note that the chosen distribution for these parameters makes only minimal difference to the results in this work.

\subsubsection{Jet kinetic powers}
\label{sec:Jet kinetic powers}

The distribution of the jet kinetic powers fuelling the extended AGN emission is informed by observations of the luminosity function of powerful FR-IIs in the 3CRR, 6CE and 7CRS samples. \citet{Willott+2001} fit the high-luminosity function for these objects at $151\rm\, MHz$ as a function of redshift. The general shape of their distribution is shown in the right-hand subplot of Figure \ref{fig:lumin_functions}, with the number density scaled to match the observed distribution at redshifts $z = 0.1$, 0.5, 1, 2 and 4. We generate our jet power distribution by converting their radio luminosity function into kinetic powers, first by using theory driven jet power--luminosity relationships; i.e. $Q\propto L_\nu^{6/7}$ \citep[e.g.][]{Willott+1999, Kaiser+2007}, noting that environment and source age introduce scatter to this relation \citep{Shabala+2013, HK+2013, Yates+2018}. The jet power corresponding to the turnover in their luminosity function is found by simulating mock catalogues (sampled every $0.2\rm\, dex$ in jet power, and including only active sources) for a range of possible turnovers and selecting the best match to the observed radio luminosity distribution. The resulting probability distribution for the kinetic jet power is given by
\begin{equation}
n(Q) = n_0 \left( \frac{Q}{Q_\star} \right)^{-1.17\alpha_\star} \exp\bigg[-\left( \frac{Q_\star}{Q} \right)^{1.17} \bigg] ,
\end{equation}
where $n_0$ is a normalisation constant, the turnover in the luminosity function occurs at $Q_\star = 10^{38.1}\rm\, W$, and the slope of the power law component of the luminosity function is $\alpha_\star = 2.27$. The shape of the probability distribution is independent of redshift, however the number density of extended AGNs increases towards higher redshifts. The radio luminosity functions generated from our mock catalogues (using the selected jet power distribution) are shown in the right-hand panel of Figure \ref{fig:lumin_functions} for a range of redshifts; these are in good agreement with the observed luminosity functions.

\subsection{X-ray luminosity function}

The mock extended radio AGN catalogue, which has been calibrated to successfully describe the observed radio properties of powerful FR-IIs, can now be used to predict their brightness at X-ray wavelengths. The integrated luminosity is calculated for active sources by summing the synchrotron emissivity from the lobe, and the bremsstrahlung radiation from both the shocked gas shell and the ambient medium. The X-ray emission from the ambient medium is included in these calculations since in practice it may be hard to disentangle the emission from the AGN and environment (except in well resolved objects), however we also derive the X-ray luminosity function including only AGN related emission for comparison with previous studies. We expect the X-ray luminosity of most extended radio galaxies to be strongly correlated with the properties of the ambient medium since the brightness of the shocked gas shell is directly related to the mass of gas swept up as the lobe expands. The level of inverse-Compton emission from the lobe, meanwhile, has a more complicated relationship with the density profile of the ambient medium. 

The accretion disk of AGNs is also often bright at X-ray wavelengths; the spectrum of the accretion disk peaks at ultraviolet wavelengths but a sizeable number of {thermal photons from the disk} are inverse-Compton upscattered to X-ray energies by the hot corona surrounding the disk. Typical X-ray luminosities from accretion-related nuclear emission are of order $10^{43}$ to $10^{45}\rm\, erg \,s^{-1}$ \citep{Mingo+2014}. Meanwhile, the highest energy shock-accelerated electrons emit synchrotron radiation at X-ray wavelengths both along the jet and at the terminal hotspot in active sources; these quickly fade in remnants due to radiative losses. The hotspots of typical FR-IIs can reach $10^{42}$ to $10^{43}\rm\, erg \,s^{-1}$ at X-ray wavelengths \citep{Harris+2000, Perlman+2010}. {The brightness of the core and hotspots is not related to the inverse-Compton upscattering of CMB photons and is thus independent of redshift, however it is hypothesised that a beamed inverse-Compton mechanism may operate in the jets \citep{Marshall+2018}}. We therefore choose not to include the (rather uncertain) core emission in our luminosity function{ whilst analyses in subsequent sections will only investigate the detection of emission spatially resolved from the core.}
{Finally, our modelling does not consider synchrotron self-Compton radiation (i.e. inverse-Compton upscattering of synchrotron photons); in Cygnus A ($z = 0.0561$), this mechanism contributes 70-80\% of non-thermal X-ray emission in the lobes \citep{deVries+2018}, however inverse-Compton upscattered CMB photons become dominant at higher redshifts ($z >0.4$ for a Cygnus A-like source).}

The X-ray luminosity function derived for a typical \emph{eRosita} survey frequency of $1\rm\, keV$ is shown in the central panel of Figure \ref{fig:lumin_functions}.  The brightness of the extended AGNs is comparable to, or at most a factor of a few brighter than (on average 30\% brighter at $z = 0.1$), typical inactive X-ray clusters in the range $10^{41} < L_{\rm X} < 10^{44}\rm\, erg\, s^{-1}$. The shape of the X-ray luminosity function also resembles that of the AGN halo mass function when including all contributors of extended X-ray emission, in particular at the lowest redshifts. As we show in Figure \ref{fig:lumin_fractions}, this is due to the dominant source of X-ray emission being bremsstrahlung radiation from the ambient medium and shocked gas shell for $z \leqslant 1$. Objects with the highest luminosities are typically the highest jet power sources, observed at ages of a few tens of Myrs when inverse-Compton radiation from the lobe begins to increase significantly, contributing in excess of 90\% of the total integrated X-ray luminosity from the AGN--host cluster system (see Figure \ref{fig:lumin_fractions}). By contrast, the X-ray luminosity function at the highest redshifts ($z > 1$) is dominated by inverse-Compton emission for the majority of luminosities {(and thus a large fraction of the extended AGN population)}. The stronger cosmic microwave background radiation at higher redshifts thus boosts the integrated luminosity well above that generated by the hot cluster gas {(on average 155\% brighter at $z = 4$)}, leading to an increased population of X-ray bright extended AGNs \citep[as previously pointed out by][]{Blundell+1999}.

\subsection{X-ray source number densities}
\label{sec:X-ray source number densities}

\begin{figure*}
\begin{center}
\includegraphics[width=0.85\textwidth,trim={95 60 115 90},clip]{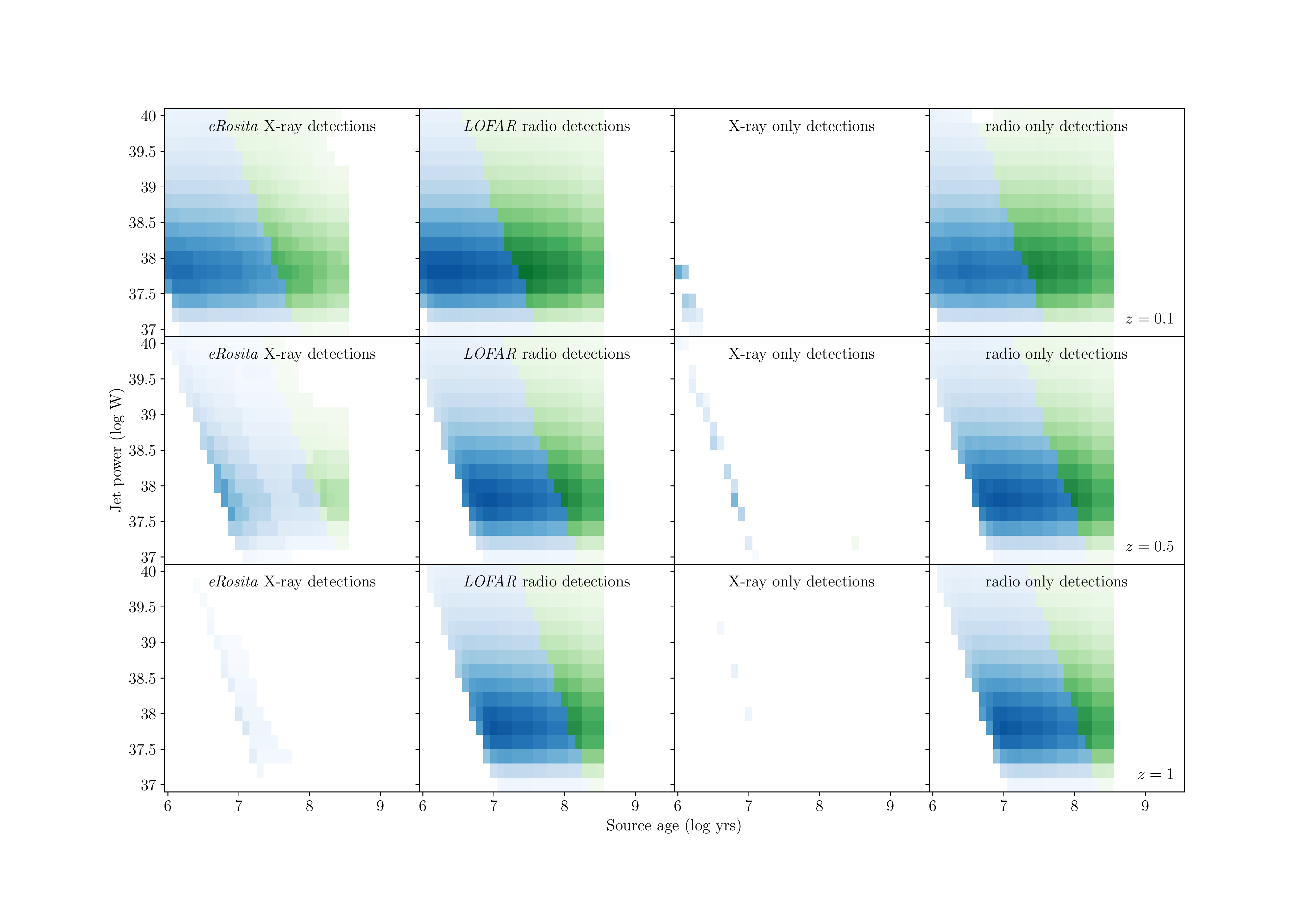}
\end{center}
\caption[]{Number density of extended AGNs in the active phase detected by an \emph{eRosita}-like X-ray survey and a \emph{LOFAR}-like radio survey as a function of their intrinsic jet kinetic power and source age. In each panel, the number density of mock objects with non-core emission detectable at X-ray and radio frequencies are shown in the {leftmost and centre-left columns}, whilst the {centre-right and rightmost columns} show the number density of objects uniquely detected at X-ray {and radio frequencies respectively}. The shading indicates the relative number of objects detected (the absolute numbers are summarised in Table \ref{tab:detections}); white shading is used for no detections with deeper colours corresponding to increasing number densities. Clearly resolved objects (i.e. those with more than five beams across the source) are shown in green and other resolved sources in blue; unresolved sources naturally are not included since non-core emission must be detected. The three {rows} show the simulated number densities for redshifts $z = 0.1$, 0.5 and 1 in the top, middle and bottom {rows respectively}.}
\label{fig:number_densities}
\end{figure*}

\begin{figure*}
\begin{center}
\includegraphics[width=0.85\textwidth,trim={95 60 115 90},clip]{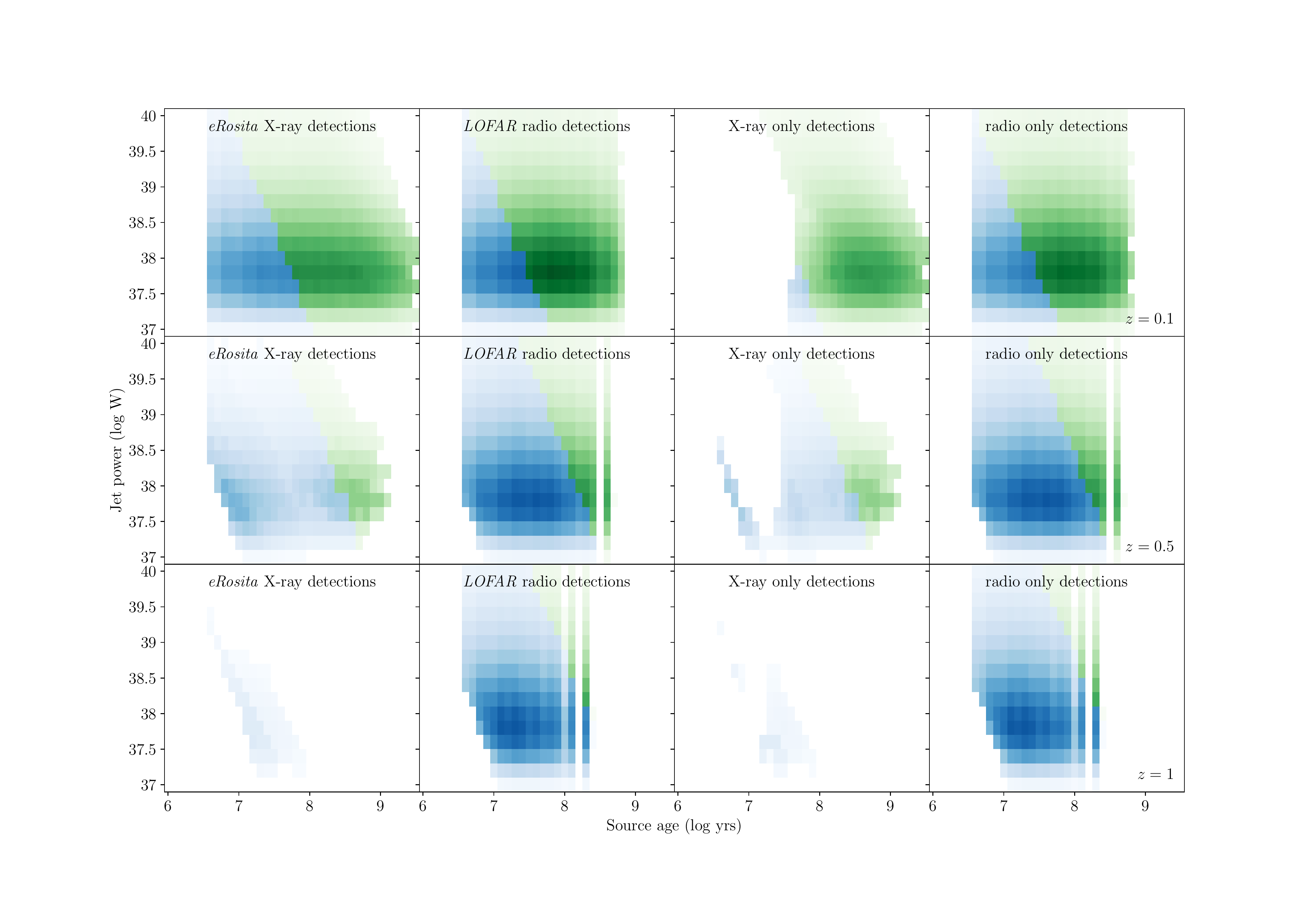}
\end{center}
\caption[]{Same as Figure \ref{fig:number_densities} but for extended AGNs of FR-II morphology in the remnant phase.}
\label{fig:remnant_densities}
\end{figure*}

We now use the mock catalogue to calculate the predicted number density of extended AGNs that would be detected using \emph{LOFAR} and \emph{eRosita}. The \emph{LOFAR Two-metre Sky Survey} (LoTSS) has a median sensitivity of $71\rm\, \mu Jy\, beam^{-1}$ at an observing frequency of 120-168\,MHz, with approximately 6\,arcsec full-width half maximum (FWHM) across the synthesised beam \citep{Shimwell+2019}. The expected $1\rm\, keV$ sensitivity of all-sky surveys using \emph{eRosita} is $14\rm\, nJy\, beam^{-1}$ for their 15\,arcsec full-width half maximum beam \citep{Merloni+2012}. We smooth the RAiSE images with a circular gaussian filter, mimicking the approximate shape of the \emph{LOFAR} and \emph{eRosita} beams. 
Extended emission separated from the core by less than the angular resolution of the survey cannot be distinguished from a bright compact core; these pixels are flagged and removed from our mock images. Simulated extended AGNs with emission in at least one beam (i.e. gaussian filtered pixel) exceeding the surface brightness sensitivity limit are assumed to be detectable by \emph{LOFAR} or \emph{eRosita} as appropriate. These sources are by definition resolved, however we make a further classification that the extended AGNs are well resolved (i.e. we can make out the shape of the source) if they have at least five beams along their length.

The number density of X-ray and radio detected active-phase, extended AGNs of FR-II morphology is shown in Figure \ref{fig:number_densities} for a range of source ages and jet kinetic powers. The overwhelming majority of low-redshift sources have extended emission resolvable from their cores in either X-ray and radio frequencies (blue or green shading); approximately half have sufficient resolution to determine the shape of their extended structures (green shading). Specifically, close to all of the active-phase FR-IIs in our catalogue are detectable using radio frequency observations (at $z \leqslant 1$), whilst 41\% can be seen at X-ray wavelengths at $z = 0.1$ and 2.6\% at $z=0.5$. At both these redshifts \emph{eRosita} can uniquely detect (i.e. no radio detection) only 0.2\% of active sources. This pattern continues out to higher redshifts where most active sources are detectable at radio frequencies whilst a rapidly decreasing fraction can be seen at X-ray wavelengths. Importantly, at $z=1$, only young and relatively compact objects are detectable using X-rays in the active phase; the source shape cannot be determined in any objects. The inverse-Compton emission arising from the overpressured lobes of these extended AGNs is related to the density of the ambient medium \citep[the dynamics of lobe expansion are described in][]{Turner+2015} and these fade rapidly as they expand into lower-density environments such as the intergalactic medium in low-mass haloes. Moreover, extended AGN emission is not detectable at redshifts $z \geqslant 2$ for the \emph{eRosita} surface brightness sensitivity limit investigated (but see Section \ref{sec:Remnant density at high-redshifts}).
The number density of extended AGNs with FR-II morphology detectable with \emph{LOFAR} or \emph{eRosita} are summarised in Table \ref{tab:detections} as a function of redshift. We scale the number densities in our catalogue based on the observed high-luminosity function of \citet{Willott+2001}.

\begin{table*}
\begin{center}
\caption[]{The number density (scaled using the \citealt{Willott+2001} sample of active phase FR-IIs) of extended AGNs detectable in the mock catalogue when including only non-core emission at \emph{eRosita} X-ray wavelengths or with a typical \emph{LOFAR} radio-frequency survey. The detected number densities are derived including either i) only active sources, ii) only remnants, or iii) any X-ray cluster excluding AGN activity, and shown for three redshift bins ($z = 0.1$, 0.5 and 1). The source number densities are calculated for an X-ray detection by \emph{eRosita}, radio detection by \emph{LOFAR}, detection at both frequencies, detection exclusively at X-ray wavelengths, and for a unique detection with the \emph{LOFAR} radio survey. The number densities in brackets are for sources with at least five \emph{eRosita} beams across their lobe; in such objects the origin of the X-ray emission can likely be determined to distinguish between hot cluster gas and AGN activity. 
}
\label{tab:detections}
\renewcommand{\arraystretch}{1.1}
\setlength{\tabcolsep}{8pt}
\begin{tabular}{ccccccc}
\hline\hline
\multirow{2}{*}{Population}&\multirow{2}{*}{Redshift}&\multicolumn{5}{c}{Detectable number density, excluding core ($\log \rm\, Mpc^{-3}$)} \\
&&\emph{eRosita}&\emph{LOFAR}&dual detection&\emph{eRosita} only&\emph{LOFAR} only
\\
\hline
\multirow{3}{*}{active phase FR-IIs}&0.1&-9.17\scriptsize{ (-10.19)}&-8.78&-9.17\scriptsize{ (-10.19)}&-11.40\scriptsize{ (-18.48)}&-9.00 \\
&0.5&-9.46\scriptsize{ (-10.69)}&-7.88&-9.49\scriptsize{ (-10.69)}&-10.61\scriptsize{ (-15.50)}&-7.89 \\
&1&-11.62\scriptsize{ (-23.86)}&-7.07&-11.63\scriptsize{ (-23.88)}&-13.12\scriptsize{ (-25.22)}&-7.07 \\
\hline
\multirow{3}{*}{remnant FR-IIs}&0.1&-9.38\scriptsize{ (-9.60)}&-9.07&-9.56\scriptsize{ (-9.95)}&-9.86\scriptsize{ (-9.87)}&-9.24 \\
&0.5&-9.59\scriptsize{ (-10.27)}&-8.13&-9.77\scriptsize{ (-11.05)}&-10.07\scriptsize{ (-10.35)}&-8.14 \\
&1&-11.80\scriptsize{ (-24.23)}&-7.54&-11.95\scriptsize{ (-25.18)}&-12.32\scriptsize{ (-24.28)}&-7.54 \\
\hline
\multirow{3}{*}{\shortstack{all X-ray clusters\\(observed mass function)}}&0.1&-4.06&--&--&-4.06&-- \\
&0.5&-10.57&--&--&-10.57&-- \\
&1&-29.70&--&--&-29.70&-- \\
\hline
\end{tabular}
\end{center}
\end{table*}

The number density of remnant FR-IIs detectable with X-ray and radio frequencies is similarly shown in Figure \ref{fig:remnant_densities}. The maximum age of remnants detectable at radio frequencies reduces quickly with increasing redshift (by more than a factor of two over the range $z=0.1$-1) due to the increasing inverse-Compton losses at high redshift. Crucially, the inverse-Compton emission remains detectable at X-ray wavelengths long after the synchrotron radiation at radio frequencies ceases. This leads to a large population of remnant extended AGNs which are detectable with \emph{eRosita} but cannot be seen at radio frequencies; these sources are typically aged between $100\rm\, Myrs$ and $1\rm\, Gyr$ (at the lowest redshifts). Specifically, \emph{eRosita} uniquely detects 14\% of remnants at $z=0.1$ (based on non-core emission), decreasing to 1.1\% and 0.002\% at redshifts $z=0.5$ and 1 respectively {(i.e. a much greater fraction of remnants can only be detected using X-ray surveys than for active sources)}. We note in Section \ref{sec:Remnant density at high-redshifts} that these number densities increase markedly for the higher redshifts with just a modest improvement in sensitivity. These exclusively X-ray detected remnants have more than five beams across the length of the lobe in 98\% of sources at redshift $z = 0.1$; this reduces to 53\% at $z = 0.5$ whilst no high-redshift ($z \geqslant 1$) remnants are observed with multiple beams across the source. The remnant source number densities are summarised in Table \ref{tab:detections}.

\subsubsection{False positive detections}
\label{sec:False positive detections}

Mock extended AGNs in our catalogue are only detected if emission from either the lobe or shocked gas shell is present above the survey surface brightness sensitivity limit. However, bremsstrahlung radiation from the brightest X-ray clusters may also be detectable using \emph{eRosita} (in a single beam) without any density enhancement from AGN activity. At redshift, $z = 0.1$, simulated X-ray clusters more massive than $10^{13.43}\rm\, M_\odot$ (derived analytically for RAiSE clusters\footnote{RAiSE simulates cluster density profiles based on the observed profiles of \citet{Vikhlinin+2006}, taking the scaling gas mass and virial radius from the semi-analytic galaxy evolution model of \citet[SAGE;][]{Croton+2016}. The gas temperature is calculated using an observed relationship with the halo mass.}) emit sufficiently high levels of bremsstrahlung radiation for their ambient medium to be detected and resolved from an X-ray nucleus. The critical cluster mass increases to $10^{14.94}\rm\, M_\odot$ at $z = 0.5$ whilst no realistic mass haloes are predicted to have their ambient medium detected (in a single beam) at redshifts $z \geqslant 1$; a small fraction of observed X-ray clusters may of course be atypically gas rich compared to our mock clusters. The number density of `false positive' detections expected due to the misclassification of cluster gas is derived by integrating the \citet{Girardi+2000} cluster mass function above the critical mass at each redshift (see Table \ref{tab:detections}).

The number density of X-ray clusters whose hot gas is detectable using \emph{eRosita} far exceeds the expected density of extended AGNs at low-redshift ($z \sim 0.1$). The presence of non-core emission at X-ray wavelengths therefore cannot be used to suggest AGN activity at these redshifts; however, 98\% of exclusively X-ray detected remnants are well-resolved (i.e. have more than five beams across their lobes) and so can potentially be classified based on the shape of their X-ray brightness distribution. The same behaviour is seen at moderate redshifts though `false positives' and actual AGN detections are expected to occur with equal likelihood. By contrast, at high-redshifts ($z \geqslant 1$) the number density of X-ray clusters detectable with \emph{eRosita} for a typical survey sensitivity is over {seventeen} orders of magnitude lower than the expected detection rate for extended AGN emission. Any non-core X-ray emission detected at high-redshifts can therefore be directly attributed to extended AGNs, and not the ambient medium. 

\subsubsection{Remnant density at high-redshifts}
\label{sec:Remnant density at high-redshifts}

\begin{figure}
\begin{center}
\includegraphics[width=0.45\textwidth,trim={10 10 40 40},clip]{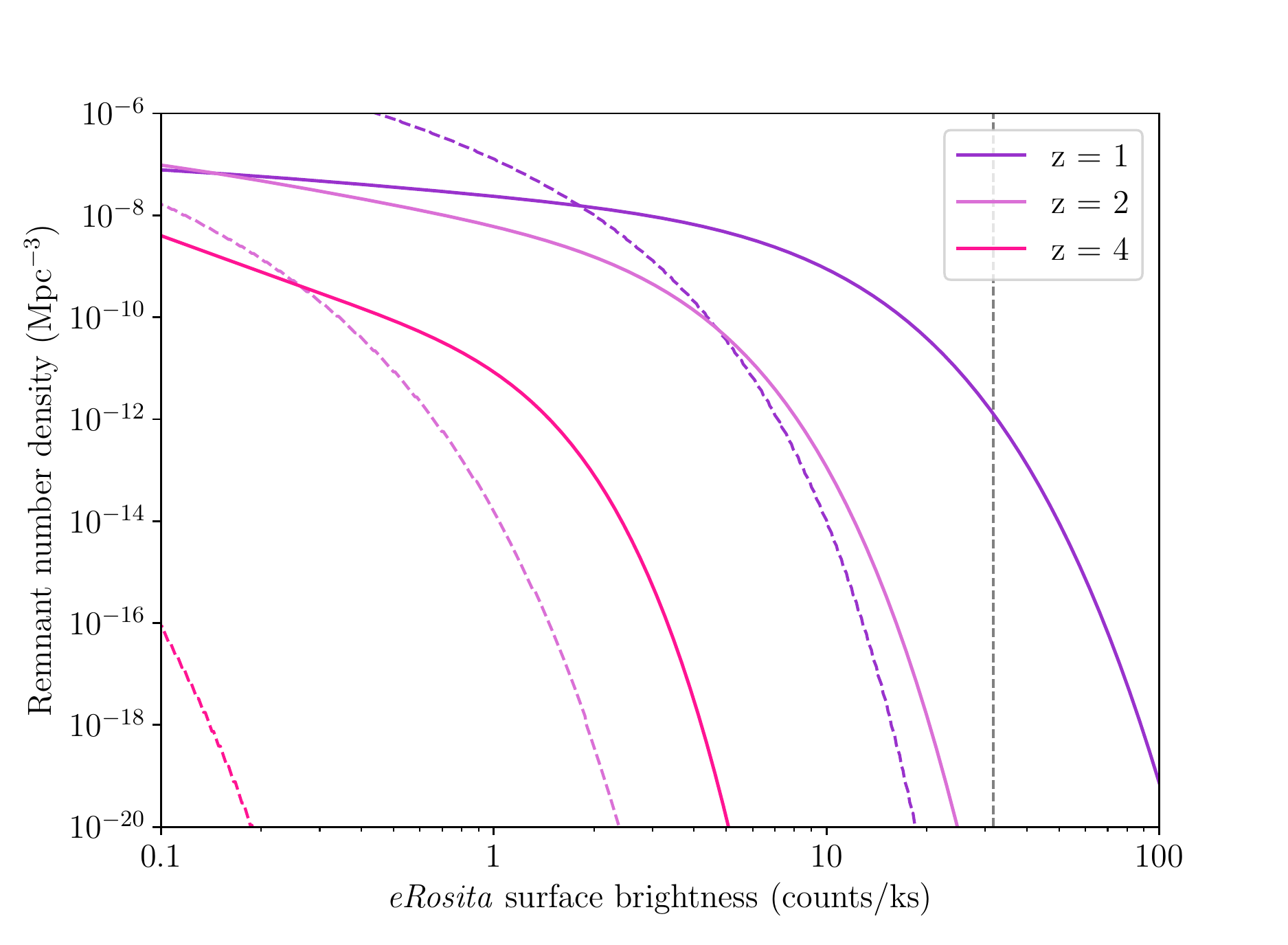}
\end{center}
\caption[]{Number density of remnant-phase extended AGNs detected by an \emph{eRosita}-like X-ray survey as a function of increasing surface brightness sensitivity. The solid lines show the theoretically detectable number of remnants (excluding core emission) at three high-redshift bins, $z=1$, 2 and 4, in increasingly red shades of purple. The dashed coloured lines show the expected false positive detection rate due to the bremsstrahlung emission from X-ray clusters at that redshift. Finally, the grey dashed vertical line shows the $14\rm\, nJy\, beam^{-1}$ surface brightness sensitivty assumed for \emph{eRosita} throughout the remainder of this work.}
\label{fig:remnant_density}
\end{figure}

The number density of extended AGNs of FR-II morphology detectable at high-redshifts using an \emph{eRosita}-like survey is investigated for a range of improved surface brightness sensitivities. In the previous section, we found none of the mock extended AGNs at redshift $z=2$ and 4 could be detected using the assumed $14\rm\, nJy\, beam^{-1}$ sensitivity; this is reduced by up to a factor of 100 down to $0.14\rm\, nJy\, beam^{-1}$ for the same resolution ($15\rm\, arcsec$ FWHM). The expected number density of remnants detectable (and mostly uniquely detectable) by this theoretical increased sensitivity \emph{eRosita} survey is shown in Figure \ref{fig:remnant_density} for redshifts $z=1$, 2 and 4 as a function of the surface brightness sensitivity. Remnants can begin to be detected at $z=2$ for a modest factor of 1.5 increase in sensitivity and at redshift $z = 4$ for an order of magnitude improvement. The rate of false positive detections from the ambient medium is also included in Figure \ref{fig:remnant_density}; the likelihood of detecting AGN emission and false positives converges at $z=1$ for a factor of 20 increase in sensitivity, but at higher redshifts the false positive rate remains small for at least a three orders of magnitude improvement in surface brightness sensitivity. 

The number density of remnants detected at $z=1$ meanwhile can increase by over {four} orders of magnitude (before converging to the false positive rate) reaching a level comparable to that found by using radio frequencies; \emph{eRosita} is therefore highly complementary to \emph{LOFAR}. The number density of radio-detected remnants does not increase with improved sensitivity as any objects emitting appreciable synchrotron emission are already detectable. Importantly, the long lasting inverse-Compton emission probes a different population of remnants that has minimal overlap with the radio-selected sample (Table \ref{tab:detections}). The number density of X-ray detected AGNs similarly coverges to the false positive rate with a comparable or greater number of detections to that obtained using radio frequencies; i.e. factor of five and 1200 more remnants at $z=2$ and 4 respectively. The detection of non-core X-ray emission at high-redshift therefore presents a viable technique to identify previous episodes of AGN activity with X-ray telescopes presently in service; this would be greatly enhanced with only a modest increase in sensitivity. Our work predicts that X-ray wavelengths are capable of detecting at least a factor of ten more remnants than at radio frequencies for redshifts $z>2.2$, increasing to a factor of 100 for redshifts $z>3.1$.

Similar results are obtained for extended AGNs in the active phase, however these are more readily observed at radio frequencies and thus are not considered likely targets for such surveys.

\section{CONCLUSIONS}
\label{sec:CONCLUSIONS}

We have extended the successful \emph{Radio AGN in Semi-analytic Environments} \citep[RAiSE][]{Turner+2015, Turner+2018a, Turner+2018} lobe expansion and evolution model to X-ray wavelengths. Specifically, this improved model considers: (1) inverse-Compton upscattering of cosmic microwave background radiation by the synchrotron-emitting electrons in the lobe; (2) the dynamics of the shocked gas shell and the associated bremsstrahlung emission from this dense gas; and (3) emission from the ambient medium surrounding the extended AGN. We construct X-ray surface brightness maps of mock extended AGNs with \citeauthor{FR+1974} type-II morphology to understand the relative importance of these radiative mechanisms; in particular, to determine what fraction of the population may be detectable (and correctly recognised as extended AGNs) at X-ray wavelengths.

X-ray and radio-frequency surface brightness maps are derived for the technical characteristics of the \emph{extended Roentgen Survey with an Imaging Telescope Array} (eRosita) and the \emph{Low-Frequency Array} (LOFAR) instruments respectively. We consider the temporal evolution of the surface brightness in the lobe (inverse-Compton or synchrotron), shocked gas shell (bremsstrahlung) and ambient medium (bremsstrahlung) along the jet axis for typical sources located at increasing redshift from $z=0.1$ to 2. The inverse-Compton emission from the lobe is dominant over the bremsstrahlung radiation from both the shocked gas shell and ambient medium in poor clusters ($10^{12.5}\rm\, M_\odot$) at all sources ages. By contrast, the shocked gas shell is initially brighter than the lobe (for between 20 and $100\rm\, Myrs$) for sources expanding into denser environments ($10^{13.5}\rm\, M_\odot$); the shocked gas shell stays brighter than the lobe for longer with either denser environments or higher jet powers. The X-ray surface brightness from the lobe and shocked gas shell remain detectable above the ambient medium {(at the $5\sigma$ level) for another 65-$115\rm\, Myrs$ after the jet switches off after $30\rm\, Myrs$ at redshifts $z \geqslant 0.5$}. By contrast, synchrotron emission ceases after only 10-$70\rm\, Myrs$ at radio frequencies. We find that although both synchrotron and inverse-Compton emission fade more rapidly at higher redshift, synchrotron radiation is much more sharply curtailed, leading to a sizable high-redshift population of remnant radio galaxies {emitting exclusively through the inverse-Compton mechanism}.

We constructed an integrated X-ray luminosity function for extended AGNs by generating a mock population of FR-IIs with jet powers, active lifetimes and host cluster environments based on observational constraints. The integrated X-ray luminosity of most extended AGNs at low redshifts ($z \leqslant 1$) is found to be strongly correlated with the properties of the ambient medium. In other words, the bremsstrahlung radiation from either the ambient medium or the shocked gas shell (whose density is strongly correlated with the host environment) is the dominant source of X-ray emission. At these low redshifts, only a small population of AGNs with the highest jet powers and moderate ages of a few tens of Myrs can have inverse-Compton emission contribute in excess of 90\% of the total integrated X-ray luminosity from the AGN--host cluster system. By contrast, the X-ray luminosity function at higher redshifts ($z > 1$) is dominated by inverse-Compton emission for the vast majority of the extended AGN population. The stronger microwave background radiation at these high redshifts boosts the lobe contribution to the integrated luminosity well above that generated by the hot cluster gas, leading to an increased population of X-ray bright extended AGNs.

We used our mock extended AGN catalogue to explore how many new objects could be detected using both existing and increased sensitivity X-ray observations. We find that most active FR-II sources at redshifts $z \leqslant 1$ can be detected at radio frequencies for the sensitivity of the \emph{LOFAR Two-metre Sky Survey} (LoTSS). However, only a small fraction of active or remnant sources can be seen at X-ray wavelengths for typical \emph{eRosita} sensitivity, when excluding core emission. No active extended AGNs are detectable at X-ray wavelengths but not at radio frequencies. By contrast, \emph{eRosita} will find 14\% of remnants at $z = 0.1$ which are not visible to \emph{LOFAR}, decreasing to 1.1\% and 0.002\% at redshifts $z = 0.5$ and 1 respectively. Meanwhile, the surface brightness of the bremsstrahlung radiation from the ambient medium of any realistic mass haloes is expected to become undetectable beyond redshift $z \geqslant 1$; i.e. any non-core X-ray detection can be attributed to extended AGN activity. We consider the effectiveness of radio-frequency and X-ray surveys at detecting remnants at these high-redshifts for greatly enhanced surface brightness sensitivity. The number density of X-ray detected remnants at redshift $z = 1$ {becomes} comparable to the number of radio-frequency detections. However, our work predicts {that} at least a factor of ten more remnants would be detected using X-ray wavelengths (compared to radio frequencies) at redshifts $z > 2.2$, increasing to a factor of 100 for redshifts $z >3.1$. Future high-sensitivity surveys using \emph{eRosita} or subsequent X-ray telescopes may therefore prove the best tool for probing the earliest generations of powerful radio galaxies.

\subparagraph{}
\noindent

{We thank an anonymous referee for helpful and constructive comments that have improved our manuscript.}



\end{document}